\documentclass[sigplan,nonacm,nocopyrightspace,10pt]{acmart}

\usepackage{hyperref}
\usepackage{booktabs} %

\usepackage{tikz}
\usetikzlibrary{arrows,automata,positioning} %
\usepackage{color}
\usepackage[normalem]{ulem}
\usepackage[english]{babel}
\usepackage[utf8]{inputenc}
\usepackage{microtype}
\usepackage{amsmath}
\usepackage{amssymb}
\usepackage{amsthm}
\usepackage[varqu]{zi4}
\usepackage{mathptmx} %
\usepackage[T1]{fontenc}
\usepackage{bbm}
\usepackage{graphicx}
\usepackage{float}
\usepackage{caption}
\usepackage{flushend}
\usepackage{subcaption,paralist,pdflscape,afterpage}
\usepackage{multirow}
\usepackage{fancyhdr}

\makeatletter
\renewcommand{\subparagraph}[1]{\vspace{2pt}\noindent {\underline{\em #1:}}}
\makeatother

\newcommand{\sonic}{{\sc {Sonic}}\xspace}
\newcommand{\tails}{{\sc {Tails}}\xspace}
\newcommand{\genesis}{{\sc Genesis}\xspace}
\newcommand{\syslong}{{\sonic}\,{\small\&}\,\tails}
\newcommand{\sys}{\sonic}
\newcommand{\metric}{IMpJ\xspace}

\usepackage{color,xspace}
\definecolor{darkgreen}{RGB}{32, 192, 32}
\definecolor{darkyellow}{RGB}{192, 140, 37}
\definecolor{darkred}{RGB}{192, 0, 0}

\renewcommand{\textrightarrow}{$\rightarrow$}

\newcommand{\arxiv}[1]{}

\usepackage{listings}
\sbox0{\small\sffamily x}
\edef\mybasewidth{\the\wd0 }
\lstdefinestyle{custompython}{%
  basicstyle=\small\sffamily,
  keywordstyle=\bfseries\color{blue!40!black},
  morecomment=[l][\itshape\color{green!40!black}]{\#},
  columns=fixed,
  basewidth=\mybasewidth,
  tabsize=4,
  keywords={def,for,if,else,in,transition,len,atomic,return,caller},
  breaklines=true}
\newcommand{\figMotivationNNs}{
  \begin{figure*}
  \centering
  \includegraphics[width=0.8\linewidth]{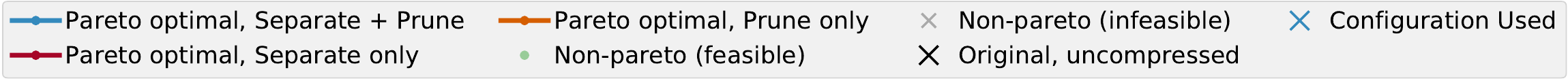}
  \vspace{0.25em}
  
  \begin{minipage}{\linewidth}
    \begin{subfigure}{0.32\linewidth}
      \includegraphics[width=\linewidth]{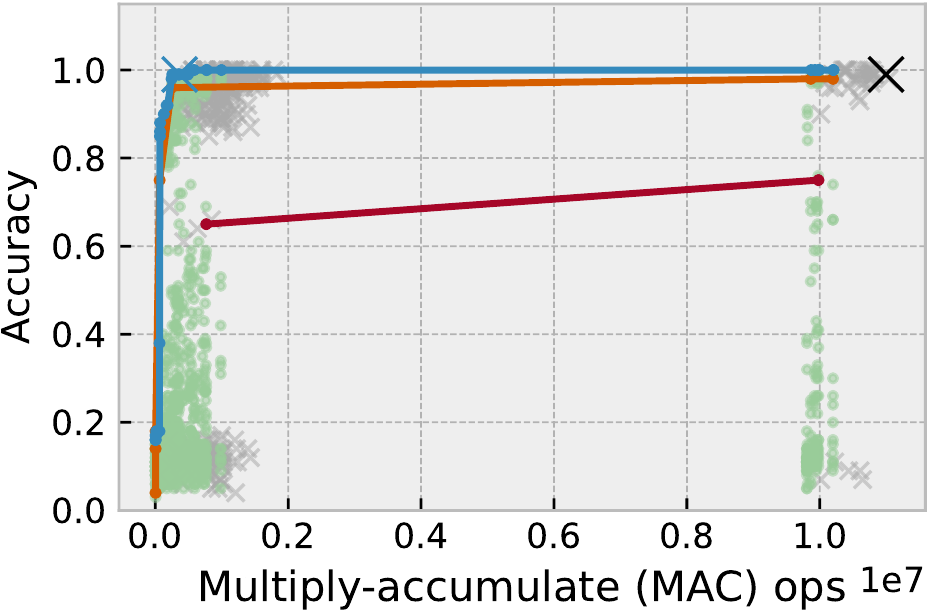}
      \caption{MNIST image recognition.}
    \end{subfigure}
    \hfill
    \begin{subfigure}{0.32\linewidth}
      \includegraphics[width=\linewidth]{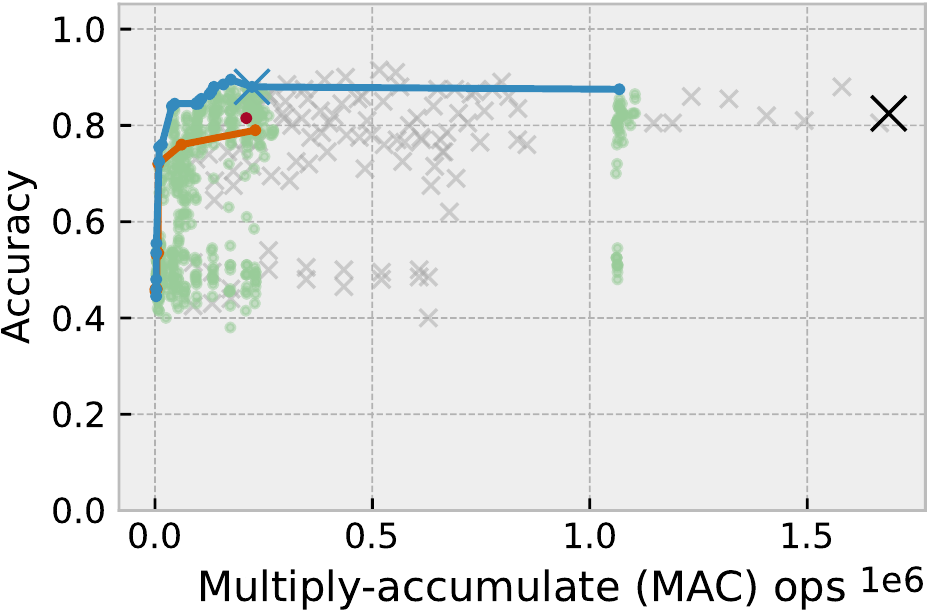}
      \caption{Human activity recognition (HAR).}
    \end{subfigure}
    \hfill
    \begin{subfigure}{0.32\linewidth}
      \includegraphics[width=\linewidth]{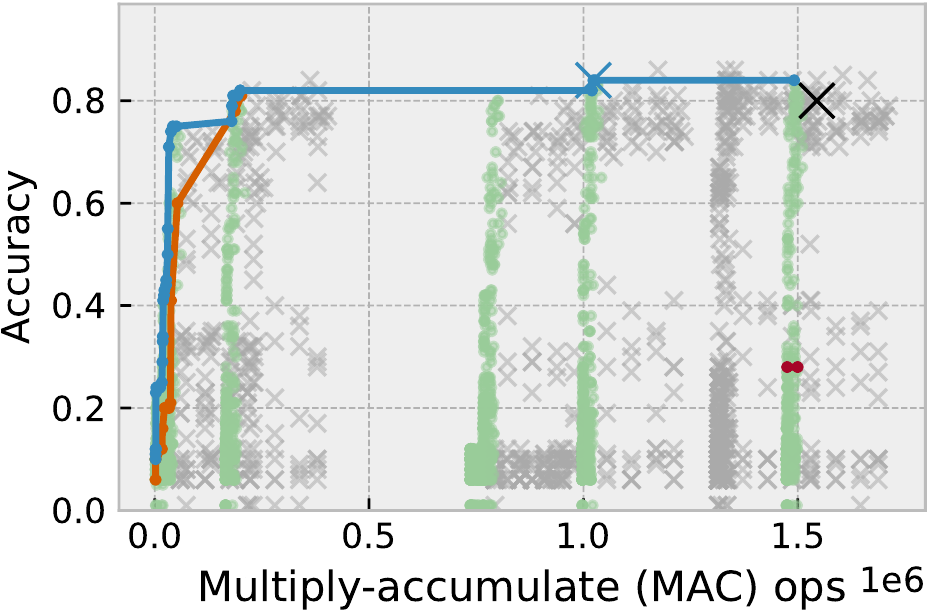}
      \caption{Google keyword spotting (OkG).}
    \end{subfigure}
    \caption{\genesis explores the inference accuracy-cost tradeoff for different neural network configurations.}
    \label{fig:genesis:train}
  \end{minipage}
  \hfill
  \begin{minipage}{\linewidth}
    \begin{subfigure}{0.32\linewidth}
      \includegraphics[width=\linewidth]{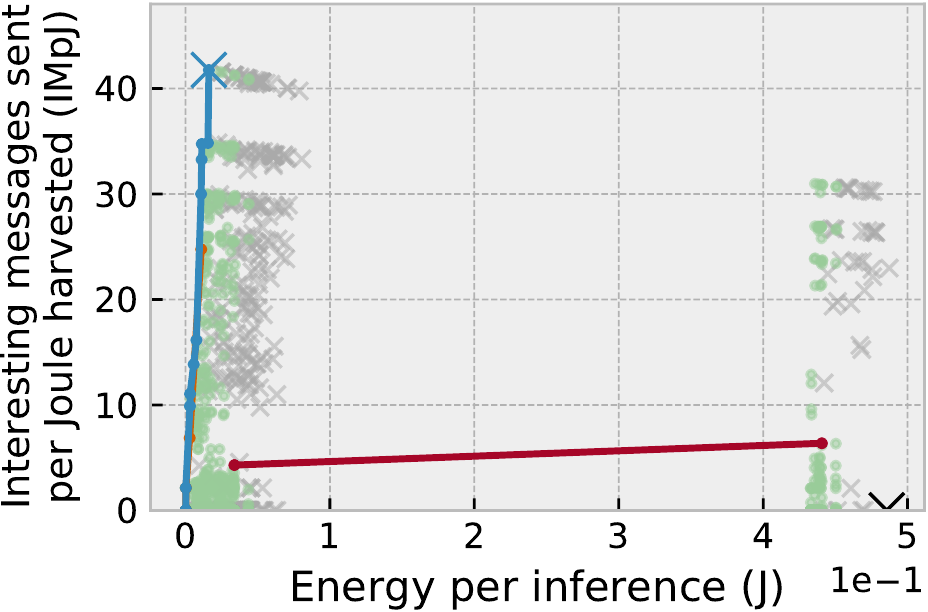}
      \caption{MNIST image recognition.}
    \end{subfigure}
    \hfill
    \begin{subfigure}{0.32\linewidth}
      \includegraphics[width=\linewidth]{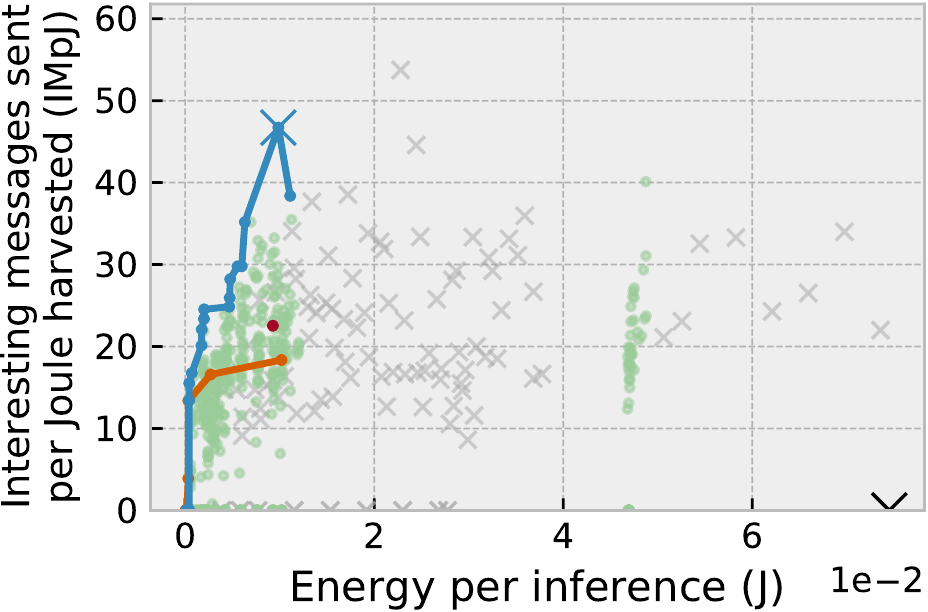}
      \caption{Human activity recognition (HAR).}
    \end{subfigure}
    \hfill
    \begin{subfigure}{0.32\linewidth}
      \includegraphics[width=\linewidth]{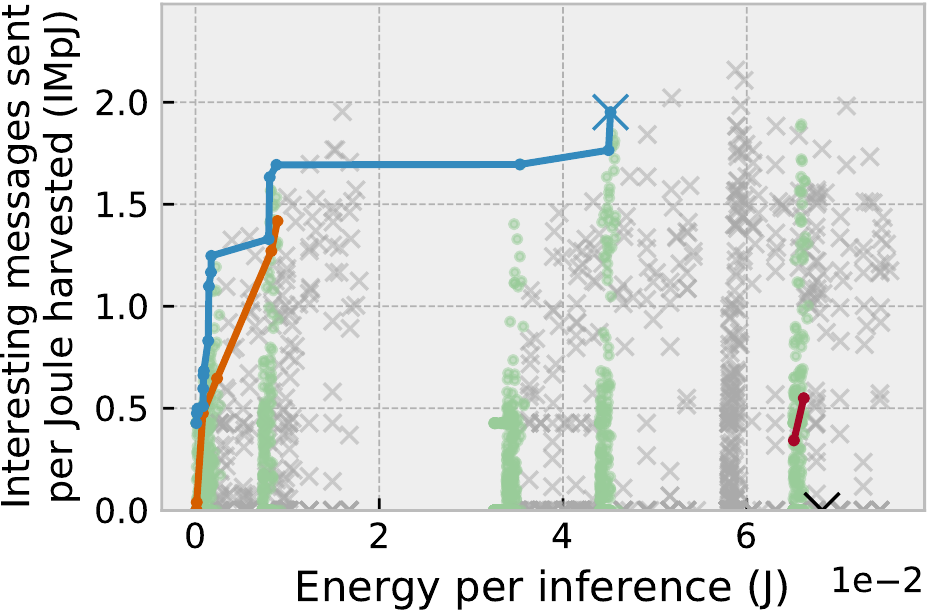}
      \caption{Google keyword spotting (OkG).}
    \end{subfigure}
    \caption{\genesis uses our end-to-end application performance
      model (\autoref{eq:impj}) to select the best feasible network configuration.}
    \label{fig:genesis:perf}
  \end{minipage}
\end{figure*}}

\newcommand{\figEvalTime}{%
\begin{figure*}
  \centering
  \includegraphics[height=0.175in]{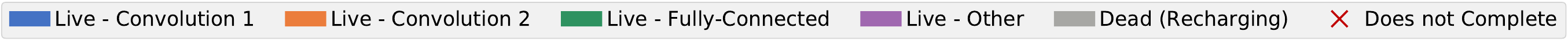}
  
  \vspace{0.25em}
  \rotatebox{90}{\small\sf Time (s)}
  \begin{subfigure}{0.32\linewidth}
    \includegraphics[width=\linewidth]{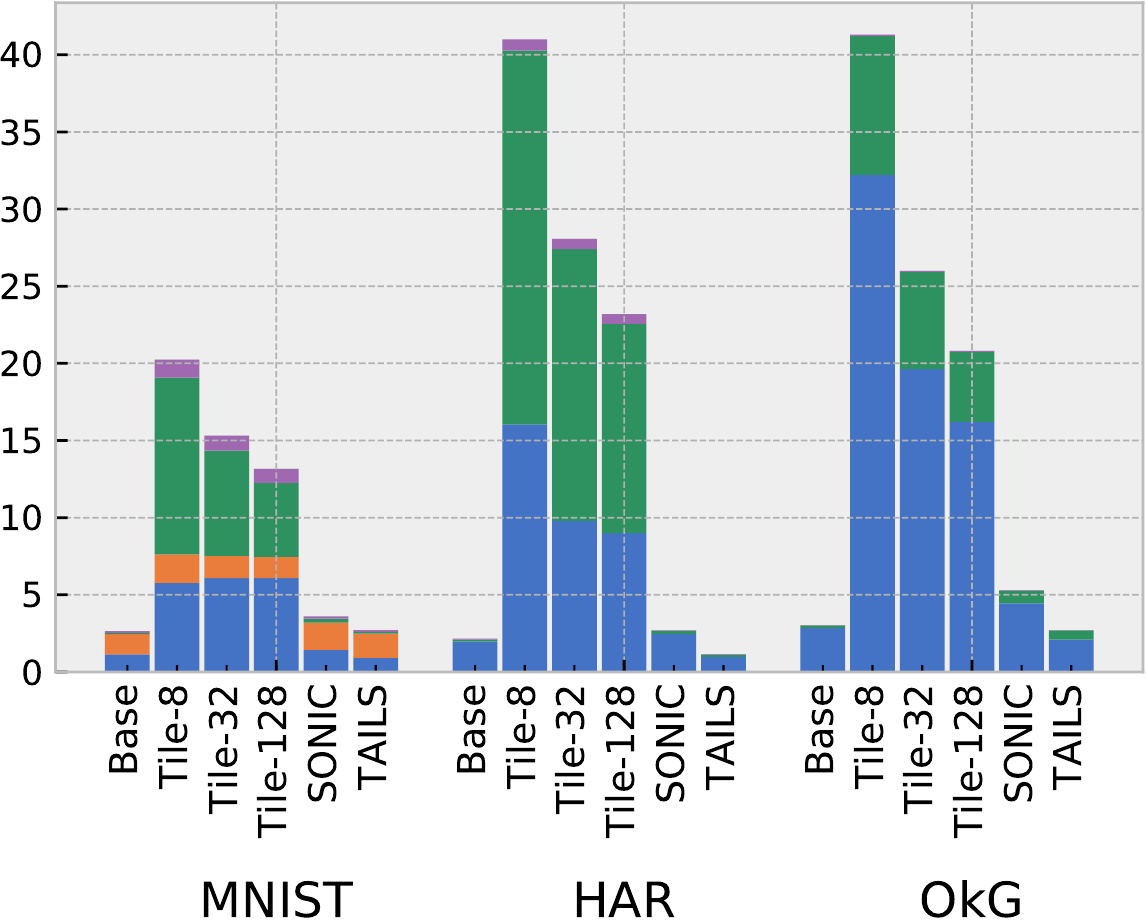}
    \caption{Continuous power.}
    \label{fig:evaluation:time:cont}
  \end{subfigure}
  \hfill
  \begin{subfigure}{0.32\linewidth}
    \includegraphics[width=\linewidth]{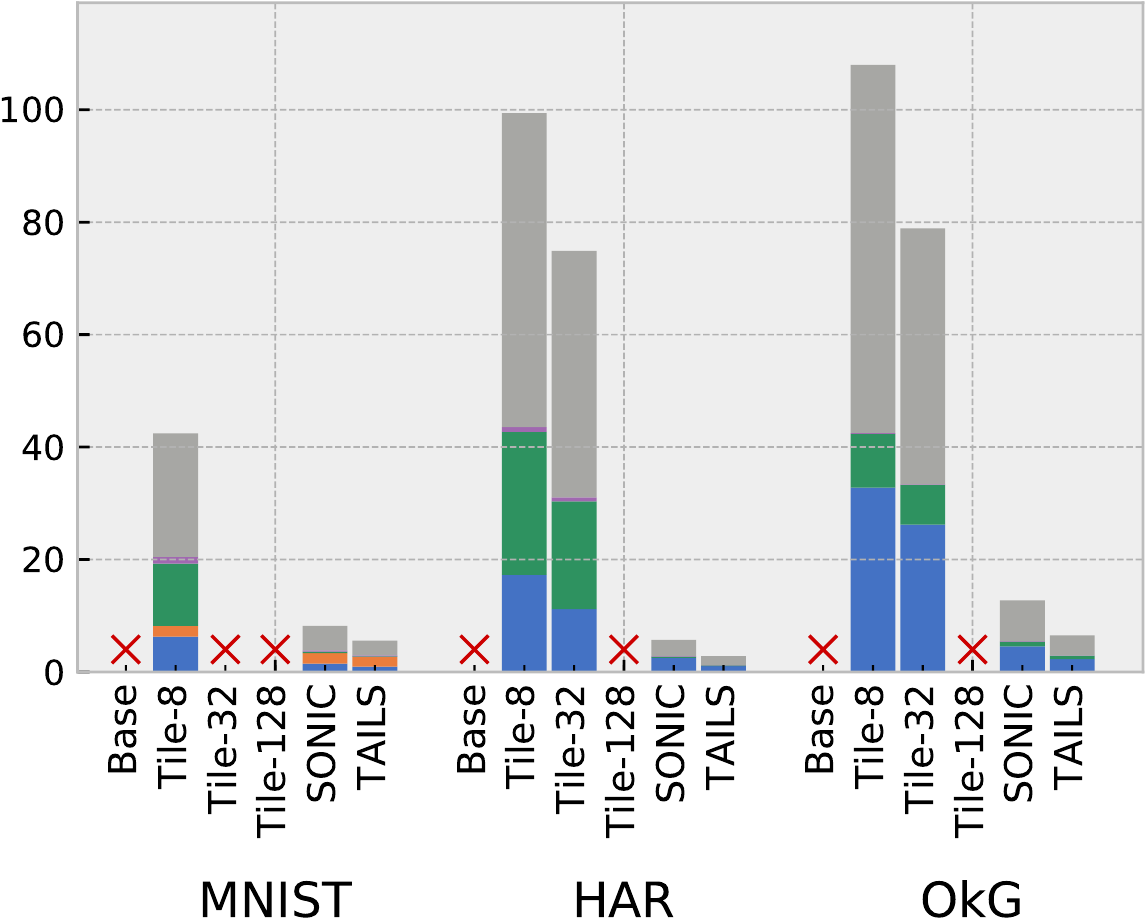}
    \caption{Intermittent power w/ 100\textmu F cap.}
    \label{fig:evaluation:time:100uf}
  \end{subfigure}
  \hfill
  \begin{subfigure}{0.32\linewidth}
    \includegraphics[width=\linewidth]{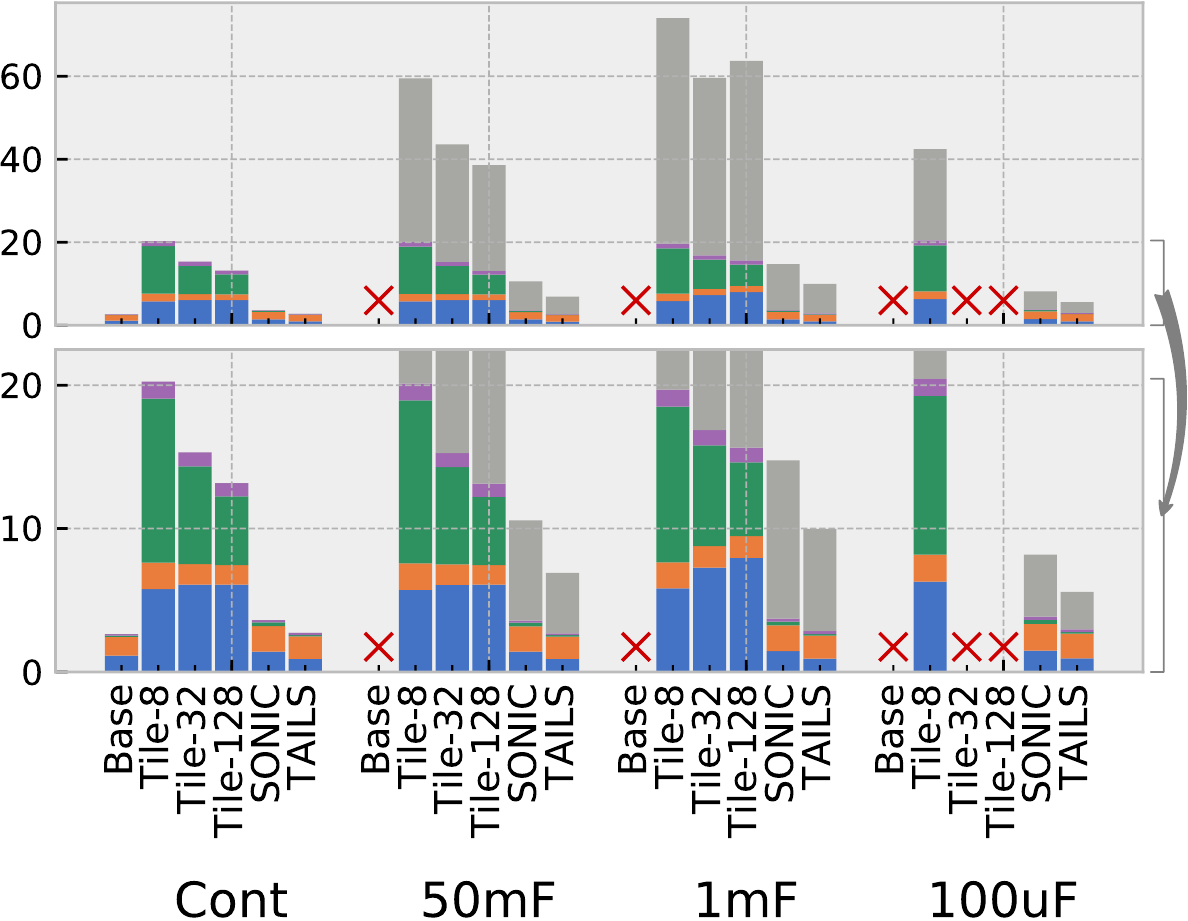}
    \caption{MNIST image recognition.}
    \label{fig:evaluation:time:mnist}
  \end{subfigure}
  
  \caption{Inference time on three neural networks.
    The na\"ive baseline is fast, but does not tolerate intermittent execution.
    Tiled implementations can ensure correct execution, but only at high cost (up to 19$\times$ slowdown) and sometimes do not complete.
    \sonic ensures correct execution and is nearly as fast as the na\"ive baseline,
    and \tails is even faster.
    (\ref{fig:evaluation:time:cont}) All three networks on continuous power,
    where \syslong add dramatically lower overheads than prior task-based systems.
    (\ref{fig:evaluation:time:100uf}) All three networks on intermittent power (100\textmu F capacitor),
    where the baseline and most tiled implementations do not complete.
    (\ref{fig:evaluation:time:mnist}) The MNIST network across all four power systems.
    \syslong always completes and has consistently good performance;
    HAR and OkG show similar results.
  }
  \label{fig:evaluation:time}
\end{figure*}}

\newcommand{\figEvalOther}{%
  \begin{figure*}
    \centering
    \vspace{-.75em}
    
    \begin{subfigure}{.32\linewidth}
      \centering
      \includegraphics[height=0.4in]{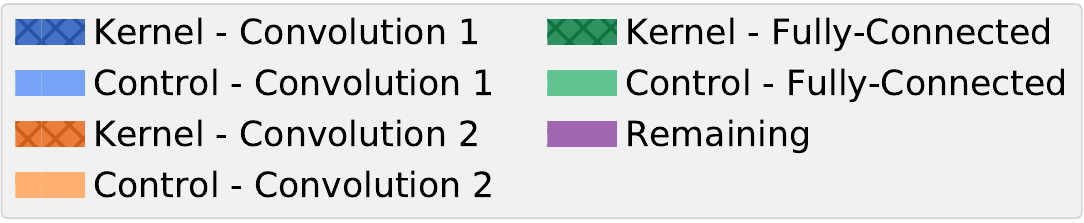}\mbox{}
    \end{subfigure}
    \hfill
    \begin{subfigure}{0.32\linewidth}
      \centering
      \hspace{8pt}
      {\includegraphics[height=0.3in]{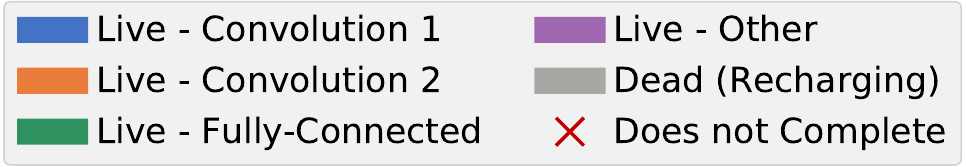}}
    \end{subfigure}
    \hfill
    \begin{subfigure}{0.32\linewidth}
      \centering
      \hspace{12pt}
      {\includegraphics[height=0.3in]{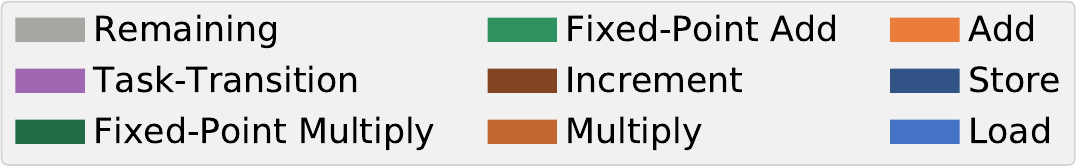}}
    \end{subfigure}

  \begin{minipage}{0.32\linewidth}
    \centering
    \rotatebox{90}{\small\sf Time (s)}  
    \hspace{-4pt}
    \begin{subfigure}{0.925\linewidth}
      {\includegraphics[width=\linewidth]{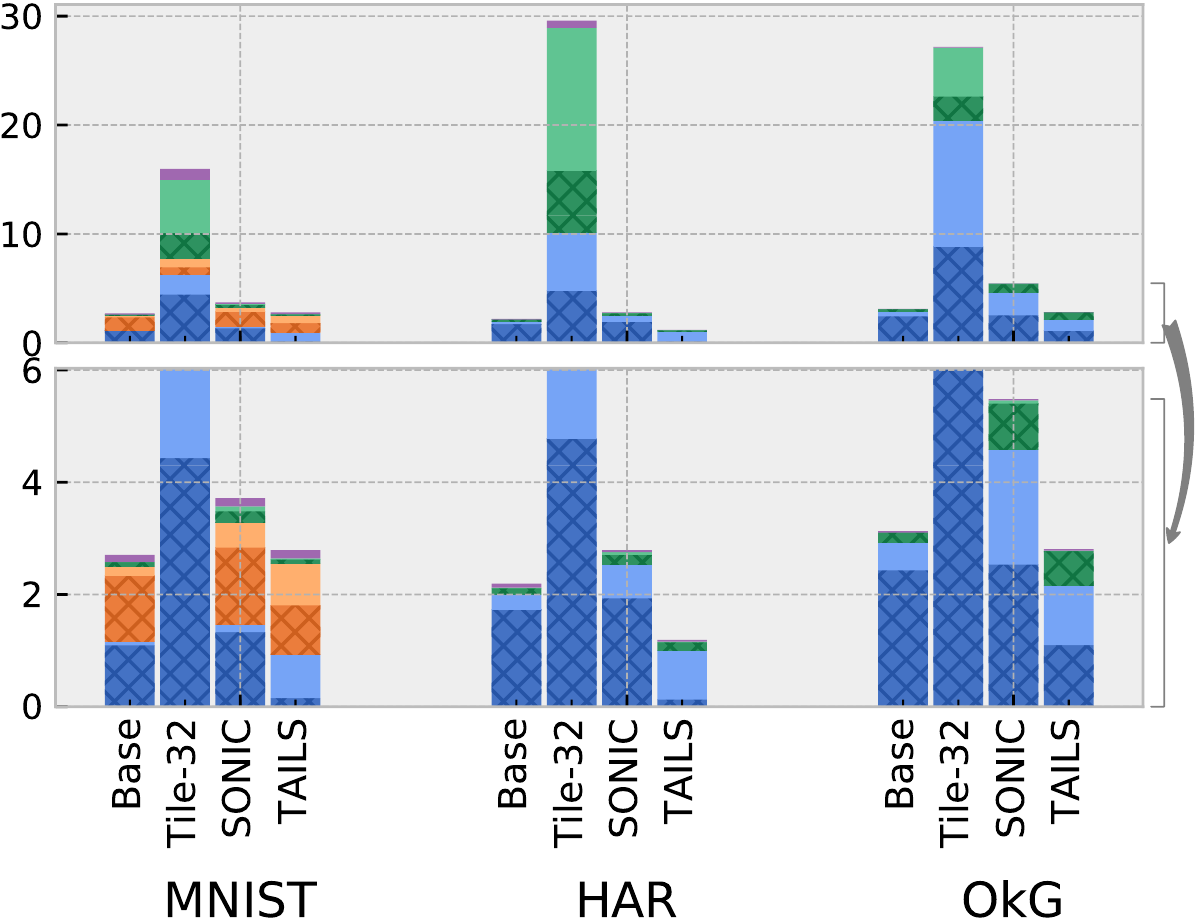}}
    \end{subfigure}
    \caption{Proportions of time spent computing the kernel of a layer.
      \syslong add small overheads over a na\"ive baseline,
      unlike prior task-based systems (Tile-32).}
    \label{fig:evaluation:time:breakdown}
  \end{minipage}
  \hfill
  \begin{minipage}{0.32\linewidth}
    \centering
    \rotatebox{90}{\small\sf Energy (mJ)}  
    \hspace{-4pt}
    \begin{subfigure}{0.925\linewidth}
      {\includegraphics[width=\linewidth]{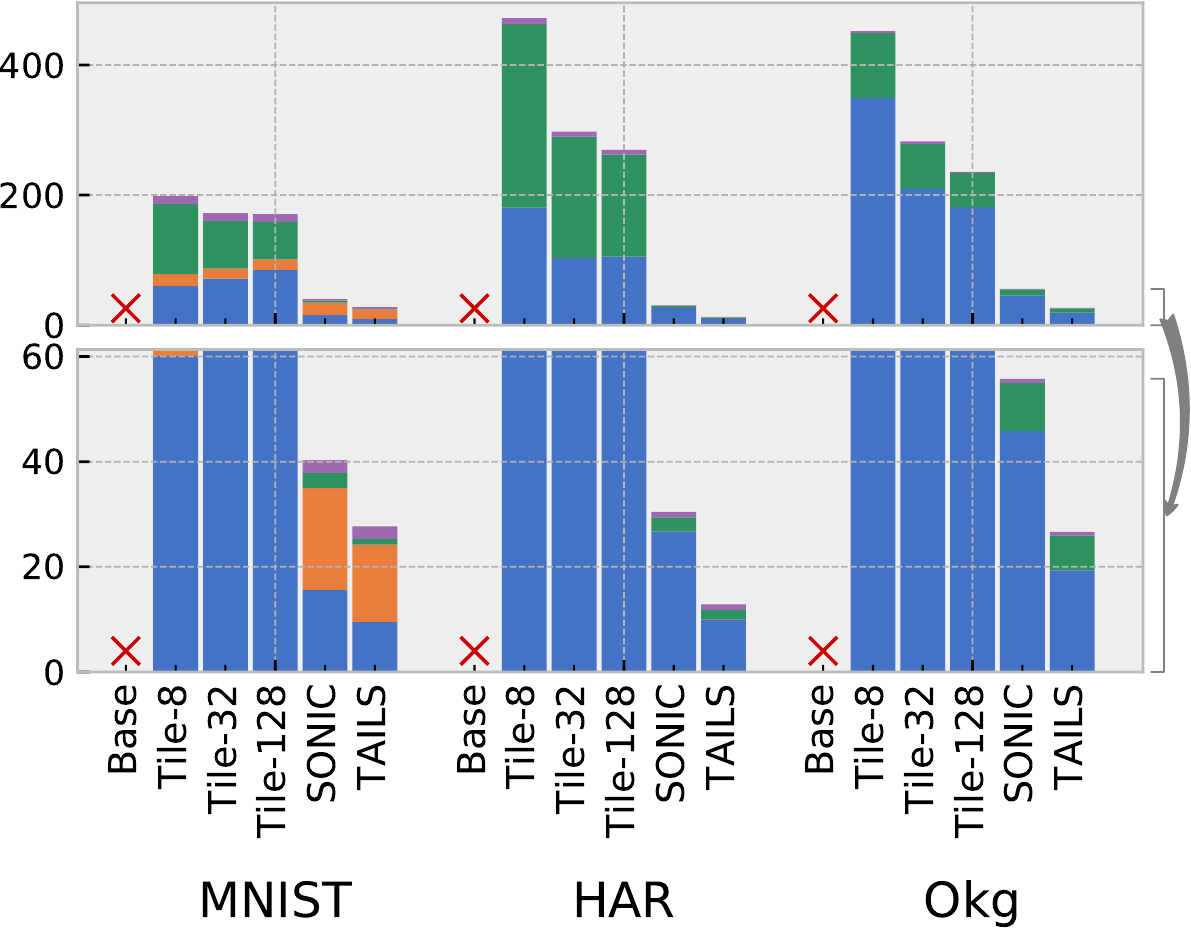}}
    \end{subfigure}
    \caption{Energy of three neural networks with a 1mF capacitor. \syslong
    require substantially less energy than the state-of-the-art.}
    \label{fig:evaluation:energy:measured}
  \end{minipage}
  \hfill
  \begin{minipage}{0.32\linewidth}
    \centering
    \rotatebox{90}{\small\sf Energy (mJ)}  
    \hspace{-4pt}
    \begin{subfigure}{0.925\linewidth}
      {\includegraphics[width=\linewidth]{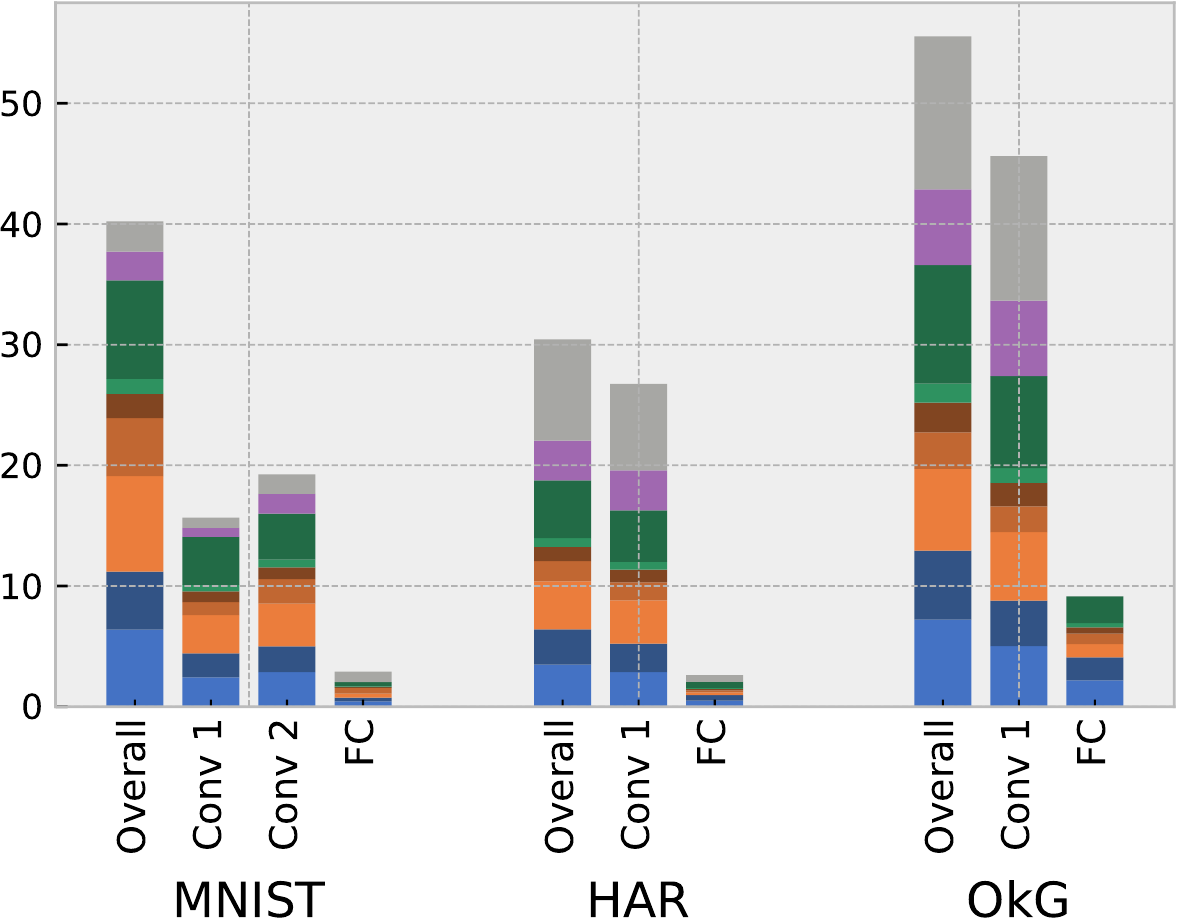}}
    \end{subfigure}
    \caption{Energy profile of \sonic broken down by operation and layer. Multiplication, control, and memory accesses represent significant overheads.}
    \label{fig:evaluation:energy:micro}
  \end{minipage}

\end{figure*}}

\begin{document}

\title{Intelligence Beyond the Edge: Inference on Intermittent Embedded Systems}

\author{Graham Gobieski}
\affiliation{%
  \institution{Carnegie Mellon University}
}
\email{gobieski@cmu.edu}

\author{Brandon Lucia}
\affiliation{%
  \institution{Carnegie Mellon University}
}
\email{blucia@cmu.edu}

\author{Nathan Beckmann}
\affiliation{%
  \institution{Carnegie Mellon University}
}
\email{beckmann@cs.cmu.edu}

\graphicspath{{figures/}}

\pagenumbering{arabic}
\pagestyle{plain}

\begin{abstract}
Energy-harvesting technology provides a promising platform for future IoT applications.
However, since communication is very expensive in these devices,
applications will require inference ``beyond the edge''
to avoid wasting precious energy on pointless communication.
We show that application performance is highly sensitive to inference accuracy.
Unfortunately, accurate inference requires large amounts of computation and memory,
and energy-harvesting systems are severely resource-constrained.
Moreover, energy-harvesting systems operate \emph{intermittently},
suffering frequent power failures that corrupt results and impede forward progress.

This paper overcomes these challenges to present the first full-scale
demonstration of DNN inference on an energy-harvesting system.
We design and implement \sonic,
an intermittence-aware software system with specialized support for DNN inference.
\sonic introduces \emph{loop continuation}, a new technique that 
dramatically reduces the cost of guaranteeing correct intermittent execution 
for loop-heavy code like DNN inference.
To build a complete system,
we further present \genesis, a tool that automatically compresses networks
to optimally balance inference accuracy and energy,
and \tails, which exploits SIMD hardware available in some microcontrollers
to improve energy efficiency.
Both \syslong guarantee correct intermittent execution without any
hand-tuning or performance loss across different power systems.
Across three neural networks on a commercially available
microcontroller, \syslong reduce inference energy by $6.9\times$ and $12.2\times$, 
respectively, over the state-of-the-art.

 \end{abstract}

\maketitle

\thispagestyle{fancy}
\setlength{\headheight}{20pt}
\fancyhf{}
\chead{\it Published in the Proceedings of the Twenty-Fourth International Conference on\\Architectural Support for Programming Languages and Operating Systems (ASPLOS'19)}

\renewcommand{\paragraph}[1]{\vspace{2pt}\noindent {\bf #1:}}
\section{Introduction}
\label{sec:intro}

The maturation of energy-harvesting technology and the recent
emergence of viable intermittent computing models creates the
opportunity to build sophisticated battery-less systems with most of
the computing, sensing, and communicating capabilities of existing
battery-powered systems.
Many future IoT applications require frequent decision making, e.g.,
when to trigger a battery-draining camera,
and these decisions must be taken locally,
as it is often impractically expensive %
to communicate with other devices.
Future IoT applications will require {\em local} inference on raw sensor data,
and their performance will be determined by inference accuracy. Using energy
numbers from recent state-of-the-art systems, we show that such local inference
can improve end-to-end application performance by 480$\times$ or more.

Recently, deep neural networks (DNNs)~\cite{alexnet, vgg, googlenet} have made
large strides in inference accuracy.
DNNs enable sophisticated inference using limited, noisy inputs, relying on
rich models learned from many examples.
Unfortunately, while DNNs are much more accurate than traditional
alternatives~\cite{gupta2017protonn, Mitchell:1997:ML:541177}, they are also
more computationally demanding.

Typical neural networks use tens of millions of weights and require
billions of compute operations~\cite{vgg,googlenet,alexnet}.
These networks target high-powered,
throughput-optimized processors like GPUs or Google's TPU, which
executes up to 9 trillion operations per second while drawing around
40 watts of power~\cite{jouppi:isca17:tpu}.
Even a small DNN (e.g., LeNet~\cite{lecun:ieee89:lenet}) has over a million
weights and millions of operations.
The most efficient DNN accelerators optimize for performance as
well as energy efficiency and consume hundreds of
mW~\cite{han:isca16:eie,chen:asplos14:diannao,du:isca15:shidiannao,chen:isca16:eyeriss}.

\paragraph{Challenges}
In stark contrast to these high-performance systems, energy-harvesting devices
use simple microcontrollers (MCUs) built for extreme low-power operation.
These MCUs systems run at low frequency (1--16 MHz) and have very small
memories (tens or hundreds of kilobytes).
Their simple architectures limit them to executing a few 
million operations per second, while consuming only 1--3mW%
---a power envelope two orders of magnitude lower than recent DNN accelerators.

DNN inference on these devices is unexplored, and several challenges
must be overcome to enable emerging IoT applications on
energy-harvesting systems built from commodity components.
Most importantly, energy-harvesting systems operate
\emph{intermittently} as power becomes available, complicating the
development of efficient, correct software.
The operating period depends on the properties of the
power system, but is short---typically around 100,000 instructions.
As a result, \emph{existing DNN inference implementations do not tolerate 
  intermittent operation}.

Recent work proposed software systems that guarantee correct
execution on intermittent power for arbitrary
programs~\cite{dino,ratchet,clank,chain,alpaca,mayfly}.
These systems add significant runtime overheads to ensure correctness,
slowing down DNN inference by on average $10\times$ in our experiments.
What these systems have missed is the opportunity to \emph{exploit the
  structure of the computation to lower the cost of guaranteeing correctness.}
This missed opportunity is especially costly for highly structured and
loop-heavy computations like DNN inference.

\paragraph{Our approach and contributions}
This paper presents the \emph{first demonstration of intermittent DNN
inference} on real-world neural networks running on a widely
available energy-harvesting system.
We make the following contributions:

\begin{compactitem}
\item We first analyze where energy is spent in an energy-harvesting system
and show that inference accuracy largely determines IoT
application performance (\autoref{sec:motivation}).
This motivates using DNNs despite their added cost over simpler but
less accurate inference techniques.
\item Building on this analysis, we present \genesis, a tool that
automatically compresses networks to maximize IoT application
performance (\autoref{sec:genesis}).
\genesis uses known compression techniques%
~\cite{nabhan1994toward, han:iclr16:deep-compression,
  chollet2016xception, bhattacharya2016sparsification};
our contribution is that \genesis optimally balances inference energy
vs.\ accuracy. %
\item We design and implement \sonic, %
  a software system for DNN inference with specialized support for intermittent execution
  (\autoref{sec:sonic}).
  To ensure correctness at low overhead,
  \sonic introduces \emph{loop continuation},
  which exploits the regular structure of DNN inference
  to selectively violate task-based abstractions from prior work~\cite{alpaca},
  allowing direct modification of non-volatile memory.
  Loop continuation is safe because
  \sonic ensures loop iterations are idempotent
  through \emph{loop-ordered buffering} (for convolutional layers) and \emph{sparse undo-logging} (for sparse fully-connected layers).
These techniques let \sonic resume from where it left off after a power failure,
{eliminating task transitions and wasted work} that plague prior task-based systems.
\item Finally, we build \tails to show how to incorporate hardware
  acceleration into \sonic (\autoref{sec:tails}). \tails uses
  hardware available in some microcontrollers to accelerate matrix
  multiplication and convolution.
  \tails automatically calibrates its parallelism
  to ensure correctness with intermittent power.

\end{compactitem}

\noindent
We evaluate \syslong on a TI MSP430
microcontroller~\cite{msp430fr5994} using an RF-energy
harvester~\cite{powercastboard,powercasttransmitter} (Secs.~\ref{sec:methodology}~\&~\ref{sec:evaluation}).
On three real-world DNNs~\cite{lecun:ieee89:lenet,okgoogle,har},
\sonic improves inference efficiency by $6.9\times$ on average
over Alpaca~\cite{alpaca}, a state-of-the-art intermittent system.
\tails exploits DMA and SIMD to further improve efficiency by $12.2\times$ on average.

We conclude with future research directions for parallel intermittent architectures
that avoid limitations of current energy-harvesting MCUs
and provide new features to support intermittence efficiently
(\autoref{sec:panic}).

\section{Background}
\label{sec:background}

Energy-harvesting devices operate using energy
extracted from their environment.  Harvested energy is not continuously
available, so an energy-harvesting device operates {\em
intermittently} as energy allows.  Prior
work showed that intermittent execution leaves memory inconsistent, compromises
progress, and suffers from non-termination conditions. 
Moreover, the typical energy-harvesting device is severely
resource-constrained, adding resource management complexity to programming.  
To motivate the contributions of \syslong,
we summarize the challenges of intermittent execution on a resource-constrained device
and describe the inefficiencies of prior intermittent execution models.

\subsection{Intermittent execution on energy-\\harvesting systems}
An energy-harvesting device operates intermittently
when harvestable power in the environment is below the device's operating
power.   To operate despite weak or periodically unavailable power, a device
slowly accumulates energy in a hardware buffer (e.g., a capacitor) and operates
when the buffer is full.
The device drains the buffer as it operates,
then it turns off and waits for the buffer to fill again.

Software executes in the {\em intermittent execution model} on an
energy-harvesting
device~\cite{mementos,dino,dewdrop,quickrecall,idetic,jerger2017ehmodel}.  In
intermittent execution, software progresses in bursts, resetting
at frequent power failures.  Existing devices~\cite{wolverine,msp430fr5994} 
mix volatile state (e.g., registers and SRAM) and non-volatile memory (e.g., FRAM). 
A power failure clears volatile state while non-volatile memory persists.
Repeated power failures impede progress~\cite{mementos}, and may leave memory
inconsistent due to partially or repeatedly applied non-volatile memory
updates~\cite{dino}.  These progress and consistency issues lead to incorrect
behavior that deviates from any continuously-powered execution~\cite{edb}.

Prior work addressed progress and memory consistency using software
checkpoints~\cite{dino,ratchet,clank}, non-volatile processors (NVPs)~\cite{nvp,ma2017incidental},
and programming models based around atomic tasks~\cite{chain,alpaca,mayfly}. 
A task-based system restarts after power loss with consistent memory at the most 
recent task or checkpoint.  We focus on task-based models because prior work 
showed that they are more efficient than checkpointing models~\cite{alpaca,chain,maeng:osdi18:chinchilla}
and because they do not rely on specialized hardware to backup architectural state after each instruction that makes
NVPs more complex and less performant.

\paragraph{Task-based intermittent execution models}
Task-based intermittent execution models avoid frequent checkpoints by
restarting from a task's start after power failure,
at which point all register and stack state must be re-initialized.
To ensure memory consistency, tasks ensure that the effect of a
partial task execution is not visible to a subsequent re-execution.
Specifically, data that are read then written (i.e., a WAR dependence) may expose the result of an
interrupted task. %
Task-based systems avoid ``the WAR problem'' with
redo-logging~\cite{alpaca} and static data duplication~\cite{chain}.

Task-based systems guarantee correct execution, but at a significant run-time cost.
Redo-logging and static duplication both increase memory and compute
in proportion to the amount of data written.
Transitioning from one task to the next takes time, so
short tasks that transition frequently suffer poor performance.
Long tasks better amortize transition costs,
but re-execute more work after a power failure.
Worse, a task that is too long faces {\em non-termination} if the energy it
requires exceeds the energy that the device can buffer.

\emph{A key challenge that we address with \syslong is ensuring correct execution of
  DNN inference while avoiding the overheads of prior task-based systems.}
We achieve this through \sonic's \emph{loop continuation},
which safely ``breaks the rules'' of existing task-based systems
by allowing WAR dependencies for loop index variables (\autoref{sec:sonic}).
This is safe because \sonic ensures that each loop iteration is idempotent.
Loop continuation yields large gains because it
effectively eliminates redo-logging, task transitions, and wasted work.

\paragraph{Resource constraints}
Intermittent systems are severely reso\-urce-constrained. In this paper we study
an intermittent system built using a TI MSP430 microcontroller (MCU), which is
the most commonly used processor in existing intermittent
systems~\cite{wisp,capybara,flicker,ufop,amulet}.
Such an MCU's frequency is typically 1--16MHz, leaving a
substantial performance gap compared to, e.g., a full-fledged, 2GHz Xeon-based
system.  
An intermittent system's MCU usually also houses all the memory available to
the system, including embedded SRAM, which is volatile, and embedded FRAM,
which is non-volatile.  
Embedded memories are small and capacity varies by device. A typical MSP430
low-power MCU includes 1--4KB of SRAM and 32--256KB of FRAM.  While
continuously powered embedded systems may interface with larger memories
via a serial bus ($i^{2}c$ or SPI), most intermittent systems do
not due to their high access energy and latency.
The typical operating power of an intermittent device is around 1mW.

\subsection{Efficient DNN inference}
Deep neural networks (DNN) are becoming the standard for inference applications
ranging from understanding speech to image recognition~\cite{alexnet,
  vgg, googlenet}.
The architecture community has responded with
accelerators that improve the performance of inference and
training and reduce power consumption.
Some architectures focus on dense computations~\cite{chen:isca16:eyeriss,
chen:asplos14:diannao, chen2014dadiannao}, others on sparse
computations~\cite{han:isca16:eie, du:isca15:shidiannao, maeri,
zhang2016cambricon}, and still others on CNN
acceleration~\cite{alwani2016fused,parashar:isca17:scnn, albericio2016cnvlutin, ding2017circnn,
ren2017sc, song2018insitu}.  
Industry has followed this trend, embracing custom silicon for
DNNs~\cite{jouppi:isca17:tpu}.

Other recent work focused on algorithmic techniques for reducing the cost of
DNN inference.
Near-zero weights can often be ``pruned'' without
losing much accuracy~\cite{nabhan1994toward,
  han:iclr16:deep-compression}.  Inference also does not need full-precision
floating-point and reducing weight
precision~\cite{han:isca16:eie,desa:isca17:sgd} reduces storage and computation
costs.
  Additional reductions in storage and
computation comes from factoring DNN
computations~\cite{nakkiran:interspeech15:compressing,
  chollet2016xception, bhattacharya2016sparsification,
  szegedy2015going, szegedy2017inception, ioffe2015batch,
  szegedy2016rethinking}.

Despite these efforts, power consumption remains orders-of-magnitude too high for energy-harvesting systems.
DNN inference consumes hundreds of milliwatts even on the most efficient accelerators~\cite{jetsontx2,chen:isca16:eyeriss,han:isca16:eie}.
Recent power-efficient DNN work from the circuits community~\cite{price2018speech,fick2017subthresholdinference} reduces power somewhat, but compromises on programmability.

More importantly, across all of these prior efforts, \emph{intermittent operation
  remains unaddressed}. It is the key problem addressed in this work.

\section{Motivation for intermittent inference}
\label{sec:motivation}

Many attractive IoT applications will be impractical without
intelligence ``beyond the edge.''
Communication is too expensive on these devices for
solutions like cloud offloading to be practical.
Instead, energy-harvesting devices must decide \emph{locally} how to
spend their energy, e.g., when to communicate sensor readings or when
to activate an expensive sensor, such as a high-resolution camera.

This section makes the case for inference on energy-harvest\-ing,
intermittently operating devices.
We show how communication dominates energy, even with
state-of-the-art low-power networking,
making cloud offloading impractical.
We analyze where energy is spent
and show that, to a first order, \emph{inference accuracy determines
  system performance},
motivating the use of DNNs in these applications.
Using this analysis we will later compare different DNN
configurations and find one that maximizes application
performance (\autoref{sec:genesis}).

\subsection{The need for inference beyond the edge}
Many applications today offload most computation to the cloud by sending input data to the cloud and waiting for a response.
Unfortunately, communication is not free.
In fact, on energy-harvesting devices,
communication costs orders-of-magnitude more energy than local computation and sensing.
These high costs mean that \emph{it is inefficient and impractical for
  energy-harvesting devices to offload inference to the edge or
  cloud}, even on today's most efficient network architectures.

For example, the recent OpenChirp network architecture lets sensors
send data over long distances with extremely low power consumption.
To send an eight-byte packet, a terrestrial sensor draws 120mA for around
800ms~\cite{dongare2017openchirp}.
Using the recent Capybara energy-harvesting power system~\cite{capybara}, such a
sensor would require a {\em 900mF} capacitor bank to send a
single eight-byte packet. 
This large capacitor array imposes an effective duty cycle on the
device, because the device must idle while charging before
it can transmit. 
A Capybara sensor node with its
2cm~$\times$~2cm solar array in direct sunlight (an optimistic setup) would take around 120 seconds
to charge a 900mF capacitor bank~\cite{capybara}. %
Hence, sending a single $28 \times 28$ image with 1B per pixel (e.g.,
one MNIST image~\cite{lecun1998mnist}) to the cloud for inference would 
take {\em over an hour}.

In contrast, our full-system \sys prototype performs inference locally
in just 10 seconds operating on weak, harvested RF energy---an
improvement of more than $360\times$. 
\syslong thus open the door to entirely new classes of inference-driven
applications on energy-harvesting devices.

\subsection{Why accuracy matters}

We now consider an example application to show how
inference accuracy determines end-to-end application performance.
This analysis motivates the use of state-of-the-art inference
techniques, namely DNNs, over less accurate but
cheaper techniques like support-vector machines.

To reach these conclusions, we employ a high-level analytical model,
where energy in the system is divided between
sensing, communication, and inference.
(Sensing includes all associated local processing, e.g., to set up the
sensor and post-process readings.)
We use local inference to filter sensor readings so that only the
``interesting'' sensor readings are communicated.
Our figure of merit is the number of interesting sensor readings
that can be sent in a fixed amount of harvested energy
(which is also a good proxy for execution time).
We denote this as \metric, or interesting messages per Joule.
Though this metric does not capture the interesting readings that are
\emph{not} communicated due to inference error (i.e., false
negatives), our analysis demonstrates the need for high accuracy,
and hence false negatives are uncommon.

This simple model captures many interesting applications of inference
beyond the edge: e.g., wildlife monitoring, disaster recovery,
wearables, military, etc.
For concreteness, we consider a wildlife-monitoring application where
sensors with small cameras are deployed across a wide area with
OpenChirp connectivity.
These sensors monitor a local population of, say, hedgehogs and send
pictures over radio when they are detected.
The goal is to capture as many images of hedgehogs as
possible, and images without have no value.

\paragraph{Baseline without inference}
Our baseline system does not support local inference, so it must
communicate every image.
Communication is expensive, so this baseline
system does not perform well.
Suppose sensing costs $E_\text{sense}$ energy,
communicating one sensor reading costs $E_\text{comm}$ energy,
and interesting events occur at a base rate of $p$
(see \autoref{tab:motivation:model}).
Then the baseline system spends $E_\text{sense} + E_\text{comm}$ energy per
event, only $p$ of which are worth communicating, and its \metric is:
\begin{equation}
  \text{Baseline} = \frac{p}{E_\text{sense} + E_\text{comm}}
\end{equation}

\begin{table}[t]
  \centering
  \resizebox{\linewidth}{!}{%
  \begin{tabular}{c p{3in}}
    \toprule
    \bf Parameter & \bf Description \\
    \midrule
    \metric & Our figure of merit, the number of ``interesting'' messages sent per Joule of harvested energy. \\
    $p$ & Base rate (probability) of ``interesting'' events. \\
    $t_p$ & True positive rate in inference. \\
    $t_n$ & True negative rate in inference. \\
    $E_\text{sense}$ & Energy cost of sensing (e.g., taking a photo). \\
    $E_\text{comm}$ & Energy cost of communicating one sensor reading. \\
    $E_\text{infer}$ & Energy cost of a inference on one sensor reading. \\
    \bottomrule
  \end{tabular}}
  \caption{Description of each parameter in our energy model.}
  \label{tab:motivation:model}
  \vspace{-1em}
\end{table}

\paragraph{Ideal}
Although impossible to build, an ideal system would communicate only
the interesting sensor readings, i.e., a fraction $p$ of all events.
Hence, its \metric is:
\begin{equation}
  \text{Ideal}
  = \frac{p}{E_\text{sense} + p \; E_\text{comm}}
\end{equation}

\paragraph{Local inference}
Finally, we consider a realistic system with local, imperfect inference.
In addition to sensing energy $E_\text{sense}$,
each sensor reading requires $E_\text{infer}$ energy to decide whether it is worth communicating.
Suppose inference has a true positive rate
of $t_p$ and a true negative rate
of $t_n$.
Since communication is very expensive, performance
suffers from incorrectly communicated,
uninteresting sensor readings at a rate of: $\left(1-p\right)\left(1-t_n\right) $.
Its \metric is:
\vspace{-1em}
\begin{align}
  \label{eq:impj}
  & \\[-0.85em]
  \text{Inference}
  &= \frac{p \; t_p}{\left(E_\text{sense} + E_\text{infer}\right) + \left(p \; t_p + \left(1-p\right)\left(1-t_n\right)\right) \; E_\text{comm}} \notag
\end{align}

\paragraph{Case study: Wildlife monitoring}
We now apply this model to the earlier wildlife monitoring example.
Hedgehogs are reclusive creatures, so ``interesting'' photos are rare,
say $p = 0.05$.
Low-power cameras allow images to be taken at low energy, e.g.,
$E_\text{sense} \approx 10$mJ~\cite{wispcam}.
As we saw above, communicating an image is expensive, taking
$E_\text{comm} \approx 23,\!000$mJ over OpenChirp~\cite{dongare2017openchirp}.
Finally, we consider two systems with local inference:
a na\"ive baseline implemented using prior task-based intermittence support (specifically Tile-8 in \autoref{sec:sonic:runtime})
and \syslong, our proposed technique.
Their inference energies are gathered from our prototype (\autoref{sec:methodology}),
taking $E_\text{infer,na\"ive} \approx 198$mJ and $E_\text{infer,\tails} \approx 26$mJ, respectively.

\begin{figure}[h]
  \centering
  \includegraphics[width=0.86\linewidth]{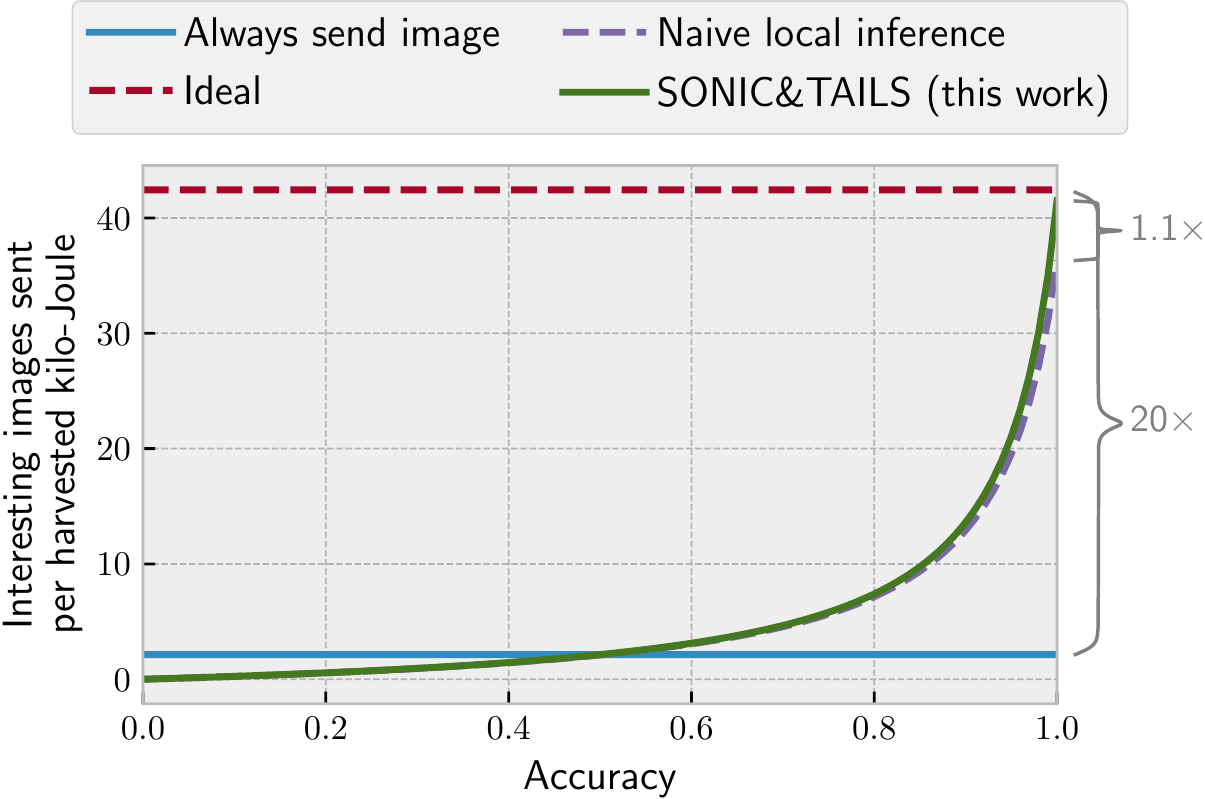}
  \caption{Inference accuracy determines end-to-end system performance
    in an example wildlife monitoring application.  Interesting events
    are rare and communication is expensive; local inference ensures
    that energy is only spent on interesting events.}
  \label{fig:motivation:images}
\end{figure}

\autoref{fig:motivation:images} shows each system's \metric after
plugging these numbers into the model.
For simplicity, the figure assumes that true positive and negative
rates are equal, termed ``accuracy''.
Since communication dominates the energy budget, local inference
enables large end-to-end benefits on the order of $\left.1\middle/p\right. = 20\times$.
However, for these gains to be realized in practice, inference must be
accurate, and the benefits quickly deteriorate as inference accuracy
declines.
Qualitatively similar results are obtained when $p$ varies, though the
magnitude of benefit changes (increasing with smaller $p$).

This system is dominated by the energy of sending results.
Inference is relatively inexpensive, so na\"ive local inference and \syslong perform similarly
(though \syslong outperforms Na\"ive by up to $14\%$).
To see the benefits of efficient inference, we must first address the system's communication bottleneck.

\paragraph{Sending only inference results}
Depending on the application, even larger end-to-end improvements are
possible by sending only the \emph{result} of inference rather than
the full sensor reading.
For instance, in this wildlife monitoring example, the
energy-harvesting device could send a single packet when hedgehogs
were detected, rather than the full image.
The effect is to significantly decrease $E_\text{comm}$
for the systems with local inference, mitigating the system's
bottleneck.
In our wildlife monitoring example, $E_\text{comm}$ decreases by $98\times$.

\begin{figure}[h]
  \centering
  \hfill\includegraphics[width=0.96\linewidth]{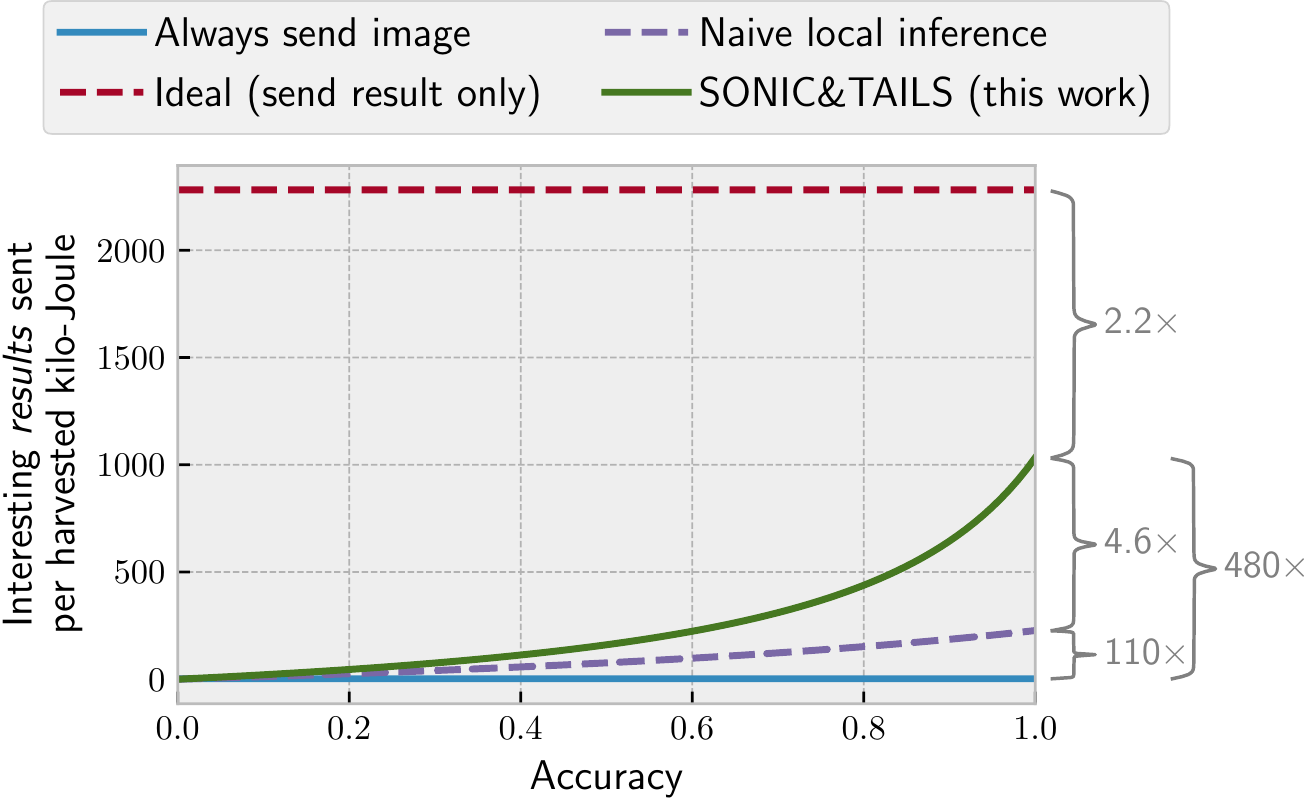}
  \caption{Local inference (i.e. Naive and \syslong) lets energy-harvesting devices communicate
    only \emph{results} of inference, enabling dramatic increases in
    end-to-end system performance.}
  \label{fig:motivation:results}
\end{figure}

\autoref{fig:motivation:results} shows end-to-end performance when
only sending inference results.
Local inference allows dramatic reductions in communication energy:
\syslong can detect and communicate $480\times$ more events than
the baseline system without local inference.
These reductions also mean that inference is a non-negligible
energy cost,
and \emph{\syslong outperform na\"ive local inference by $4.6\times$.}
Finally, the gap between Ideal and \syslong is $2.2\times$.
This gap is difficult to close further on current hardware;
we discuss ways to address it in \autoref{sec:panic}.

\section{System overview}
\label{sec:overview}

\begin{figure}[t]
\centering
\includegraphics[width=0.95\linewidth]{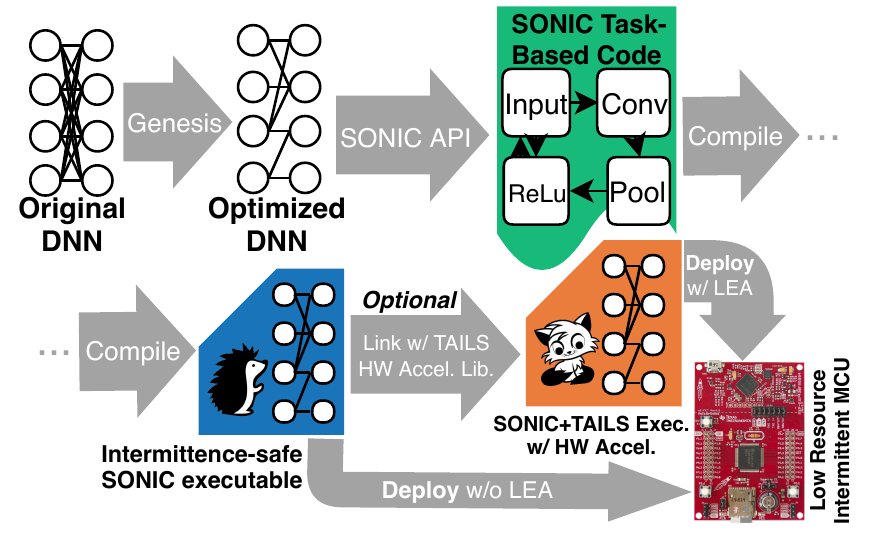}
\caption{\label{fig:overview} Overview of implementing a DNN application using \syslong.
  \genesis first compresses the network to optimize interesting messages sent per Joule (\metric).
  \syslong then ensure correct intermittent execution at high performance~\cite{Gobieski2018IntermittentDN}.
}
\vspace{-1em}
\end{figure}

This paper describes the first system for performing DNN inference
efficiently on intermittently-operating, energy-harvest\-ing
devices. \autoref{fig:overview} shows the new system components in this work
and how they produce an efficient, intermittence-safe executable starting from
a high-level DNN model description.  There are three main components to the
system: \genesis, \sonic, and \tails.

\genesis ({\underline g}enerating {\underline e}nergy-aware {\underline n}etworks for {\underline
  e}fficien{\underline s}y on {\underline i}ntermittent {\underline s}ystems) is a tool that automatically
optimizes a DNN, starting from
a programmer's high-level description of the network.  \genesis attempts to compress each layer of the network using well-known separation and pruning techniques.
\genesis's goal is
to \emph{find a network that optimizes \metric}
while meeting resource
constraints.  As \autoref{fig:overview} shows, \genesis's input is
a network description and its output is an optimally compressed network. \autoref{sec:genesis} describes \genesis.

\sonic ({\underline s}oftware-{\underline o}nly {\underline n}eural {\underline i}ntermittent {\underline
c}omputing) is an intermittence-safe,
task-based API and runtime system that
includes specialized support for DNN inference
that \emph{safely ``breaks the rules'' of existing task-based systems
to improve performance}.
\sonic is compatible with existing task-based
frameworks~\cite{chain,alpaca}, allowing seamless integration into larger applications.
\autoref{sec:sonic} describes \sonic in detail.

\tails ({\underline t}ile-{\underline a}ccelerated {\underline i}ntermittent {\underline L}EA {\underline
s}upport) is an alternative to the \sonic runtime library that
leverages hardware vector acceleration, specifically targeting the TI Low
Energy Accelerator (LEA)~\cite{lea}.  To use \tails, the
programmer need only link their compiled binary to the \tails-enabled
runtime system.  This runtime includes all of \sonic's optimizations and a
suite of hardware-accelerated vector operations, such as convolutions. 
\autoref{sec:tails} describes \tails in detail.

Starting with a high-level network description, a
programmer can use \genesis, \sonic, and \tails to build an efficient,
intermittent DNN-enabled application that meets resource constraints, is robust
to intermittent operation, and leverages widely available hardware acceleration.
Our code and datasets can be found at: \textsf{\href{https://github.com/CMUAbstract/SONIC}{https://github.com/CMUAbstract/SONIC}}. 
\figMotivationNNs

\section{Optimal DNN compression with \genesis}
\label{sec:genesis}

The first challenge to overcome in \syslong is fitting neural networks into the
resource constraints of energy-harvesting systems.
In particular, the limited memory capacity of current microcontrollers
imposes a hard constraint on networks.
We have developed a tool called \genesis
that automatically
explores different configurations of a baseline neural network,
applying separation and pruning techniques (\autoref{sec:background}) to reduce the network's resource requirements.
\genesis improves upon these known techniques by
optimally balancing inference energy and true positive/negative rates to maximize \metric,
building on the the model in \autoref{sec:motivation}.

\subsection{Neural networks considered in this paper}

This paper considers three networks, summarized in \autoref{tab:nns}.
To represent image-based applications (e.g., wildlife monitoring and
disaster recovery), we consider MNIST~\cite{lecun1998mnist}. We consider
MNIST instead of ImageNet because ImageNet's large images 
do not fit in a resource-constrained device's memory.
To represent wearable applications, we consider human activity recognition
(HAR). HAR classifies activities using accelerometer data~\cite{har}.
To represent audio applications, we consider Google keyword
spotting (OkG)~\cite{okgoogle}, which classifies words in audio snippets.

We also evaluated binary neural networks and several SVM models
and found that they perform poorly on current energy-harvesting MCUs.
A 99\%-accurate binary network for MNIST required 4.4MB of weights~\cite{courbariaux2016binarized},
exceeding the device's scant memory, and compressing this to 360KB lost nearly 10\% accuracy~\cite{binarynetgithub}.
Likewise, no SVM model that fit on the device was competitive with 
the DNN models~\cite{lecun1998gradient}: measured by \metric, SVM under-performed 
by 2$\times$ on MNIST and by 8$\times$ on HAR, and we could not find an SVM model 
for OkG that performed anywhere close to the DNN.

\subsection{Fitting networks on energy-harvesting systems}

\genesis evaluates many compressed configurations of a 
network and builds a Pareto frontier.
Compression has trade-offs in four dimensions, difficult to capture with a pareto curve;
these include true negative rate, true positive rate, memory size (i.e.,
parameters), and compute/energy (i.e., operations).
Fully-connected layers typically dominate memory, whereas convolutional layers
dominate compute. \genesis compresses both.

\genesis compresses each layer using two known techniques: separation and pruning.
Separation (or rank decomposition) splits an $m \times n$
fully-connected layer into two $m \times k$ and $k \times n$ matrix
multiplications, or an $m \times n \times k$ convolutional filter into three $m
\times 1 \times 1$, $1 \times n \times 1$, and $1 \times 1 \times k$, 
filters~\cite{chollet2016xception, bhattacharya2016sparsification}.
\genesis separates layers using the Tucker tensor decomposition,
using the high-order orthogonal iteration algorithm~\cite{tucker1966some, de2000best, de2000multilinear}.
Pruning involves removing parameters below a given threshold, since
they have small impact on results~\cite{nabhan1994toward, han:iclr16:deep-compression}.

\genesis sweeps parameters for both separation and pruning across each layer of
the network, re-training the network after compression to improve
accuracy.
\genesis relies on the Ray Tune black box optimizer with the Median
Stopping Rule to explore the configuration space~\cite{golovin2017google, moritz2017ray}.
\autoref{fig:genesis:train} shows the results for the networks in
\autoref{tab:nns}.
Each marker on the figure represents one compressed configuration,
shown by inference accuracy on the $y$-axis and inference energy on
the $x$-axis.
Feasible configurations (i.e., ones that fit in our device's 
small memory; see \autoref{sec:methodology}) are shown as green
circles and infeasible configurations are grey $\times$s.
Note that the original configuration (large $\times$) is infeasible for all three networks,
meaning that they cannot be na\"ively ported to the device because their parameters would not fit in memory.

\autoref{fig:genesis:train} also shows the Pareto frontier for each
compression technique.
Generally, pruning is more effective than separation, but the
techniques are complementary.

\begin{table}[t]
  \centering
  \resizebox{\linewidth}{!}{
  \begin{tabular}{ccccccc}
    \toprule
    \bf Network & \bf Layer & \bf Uncompressed & \bf Compression & \bf Compressed &
    \bf Compression & \bf Accuracy \\
    & & \bf Size & \bf Technique & \bf Size & &\\
    \midrule
    \multirow{4}{*}{\parbox{0.7in}{Image classification (MNIST)}}
    & Conv & $20\times1\times5\times5$& HOOI & 3$\times$1D Conv & 11.4$\times$ &
    \multirow{4}{*}{$99.00\%$}\\
    & Conv & $100\times20\times5\times5$ & Pruning& 1253 & $39.9\times$ &\\
    & FC & $200\times1600$ & Pruning, SVD & 5456 & $ 109\times$ &\\
    & FC & $500\times200$ & Pruning, SVD & 1892 & --- &\\
    & FC & $10\times500$ & --- & --- & --- &\\
    \midrule
    \multirow{4}{*}{\parbox{0.7in}{Human activity recognition (HAR)}}
    & Conv & $98\times3\times1\times12$ & HOOI & 3$\times$1D Conv&
    $2.25\times$
    & \multirow{4}{*}{$88.0\%$}\\
    & FC & $192\times2450$ & Pruning, SVD & 10804 & \multirow{2}{*}{$58.1\times$}&\\
    & FC & $256\times192$ & Pruning, SVD & --- & --- & \\
    & FC & $6\times256$ & --- & --- & --- & \\
    \midrule
    \multirow{4}{*}{\parbox{0.7in}{Google keyword spotting (OkG)}}
    & Conv %
           & $186\times1\times98\times8$ & HOOI, Pruning & 3$\times$1D Conv & 7.3x&
           \multirow{4}{*}{$84.0\%$}\\
    & FC & $96\times1674$ & Pruning, SVD & 16362 & $11.8\times$&\\
    & FC & $128\times96$ & Pruning, SVD & 2070 & --- &\\
    & FC & $32\times 128$ & SVD & 4096 & 2$\times$ &\\
    & FC & $128\times32$ & SVD & 4096 & --- &\\
    & FC & $128\times12$  & --- & --- & --- &\\
    \bottomrule
  \end{tabular}}
  \caption{Neural networks used in this paper.}
  \label{tab:nns}
  \vspace{-2.3em}
\end{table}

\subsection{Choosing a neural network configuration}

\genesis estimates a configuration's \metric using the model
from \autoref{sec:motivation}, specifically \autoref{eq:impj}.
The user specifies $E_\text{sense}$ and $E_\text{comm}$ for their
application as well as per-compute-operation energy cost.
From these parameters, \genesis estimates $E_\text{infer}$ for each
configuration, and uses the inference accuracy from the prior training step to
estimate application performance.
The user can specify which class in the training set is
``interesting,'' letting \genesis compute true positive $t_p$ and negative $t_n$ rates for the
specific application.

\autoref{fig:genesis:perf} shows the results by mapping each point in
\autoref{fig:genesis:train} through the model.
For these results, we use $E_\text{sense}$ from
\autoref{sec:motivation}, per-operation energy from our \syslong
prototype in \autoref{sec:methodology}, and estimate $E_\text{comm}$
from input size assuming OpenChirp networking~\cite{dongare2017openchirp}.

\genesis chooses the feasible configuration that maximizes
estimated end-to-end performance (i.e., \metric).
\autoref{fig:genesis:perf} shows that this choice is non-trivial.
True positive, true negative, and inference energy affect end-to-end application
performance in ways that are difficult to predict. Simply choosing the most accurate
configuration, as the twisty blue curve suggests in \autoref{fig:genesis:perf}, 
is insufficient since it may waste too much energy or underperform other 
configurations on true positive or true negative rates.

\section{Efficient intermittent inference with \sonic}
\label{sec:sonic}

\sonic is the first software system optimized for inference on
resource-constrained, intermittently operating devices.
\sonic supports operations common to most DNN computations,
exposing them to the programmer through a simple API.
\sonic's functionality is implemented as a group of {\em tasks}
supported by the \sonic runtime system, which is a modified version of the
Alpaca runtime system~\cite{alpaca}.
These tasks implement DNN functionality, and the \sonic runtime system 
guarantees correct intermittent operation.

Specializing intermittence support for DNN inference yields large benefits.
Prior task-based intermittent execution models~\cite{alpaca,chain} can degrade performance by up to 19$\times$
and by 10$\times$ on average
(\autoref{sec:evaluation}).
\sonic dramatically reduces these overheads to just 25\%--75\% over
a standard baseline of DNN inference that does not tolerate intermittent operation.

\sonic achieves these gains by eliminating the three major sources of overhead in prior task-based systems:
redo-logging, task transitions, and wasted work (\autoref{sec:background}).
Our key technique is \emph{loop continuation},
which selectively violates the task abstraction
for loop index variables.
Loop continuation lets \sonic directly modify loop indices without frequent and expensive saving and restoring.
By writing loop indices directly to non-volatile memory,
\sonic checkpoints its progress after each loop iteration,
eliminating expensive task transitions
and wasting work upon power failure.

Loop continuation is safe because \sonic
ensures that each loop iteration is idempotent.
\sonic ensures idempotence 
in convolutional and fully-connected layers
through \emph{loop-ordered buffering} and \emph{sparse undo-logging}.
These two techniques ensure idempotence without statically privatizing or dynamically checkpointing data,
avoiding the overheads imposed by prior task-based systems.

\subsection{The \sonic API}

The \sonic API lets the programmer describe a DNN's structure through 
common linear algebra primitives.
Just as a programmer chains tasks together in a task-based intermittent
programming model~\cite{chain,alpaca,mayfly}, the programmer chains \sonic's
tasks together to represent the control and data flow of a DNN inference
pipeline. 
\sonic's API exposes functionality that the programmer invokes like
any other task in their program (specifically, a {\em modular task
  group}~\cite{alpaca,chain}).
Though \sonic ``breaks the rules'' of a typical task-based
intermittent system, the programmer does not need to reason about these
differences when they are writing a program using the \sonic API.
The program-level behavioral guarantee that \sonic provides is the
same as the one underlying other task-based intermittent execution
models: a \sonic task will execute atomically despite power
interruptions by ensuring that repeated, interrupted attempts to
execute are idempotent.

\subsection{The \sonic runtime implementation}
\label{sec:sonic:runtime}

DNN inference is dominated by loops within each layer of the neural network.
\sonic optimizes DNN inference by ensuring that these loops execute correctly
on intermittent power while adding much less overhead than prior task-based systems.

\paragraph{Loops in task-based systems}
A typical task-based intermittent system sees two kinds of loops: {\em short
loops} and {\em long loops}.  All iterations of a {\em short loop} fit in a
single task and will complete without consuming more energy than the device can
buffer.  A short loop maintains control state in volatile memory and these
variables clear on power failure. When power resumes, the task restarts and
completes.  Data manipulated by a short loop are usually non-volatile (i.e.,
``task-shared''~\cite{alpaca}) and if read and updated, they must be backed up
(either statically or dynamically) to ensure they remain consistent.  The
problem with short loops is that they always restart from the beginning,
wastefully repeating loop work that was already done. 
In contrast, a {\em long loop} with many iterations does not fit in a single
task; a long loop demands more energy than the device can buffer and may never
terminate. A programmer must split loop iterations across tasks, requiring a
task transition on each iteration and requiring control state and data to be
non-volatile and backed up.  
The problem with long loops is that may not terminate and, when
split across tasks, impose hefty privatization and task transition overheads.

\begin{figure}[t]
\centering
\includegraphics[width=\linewidth]{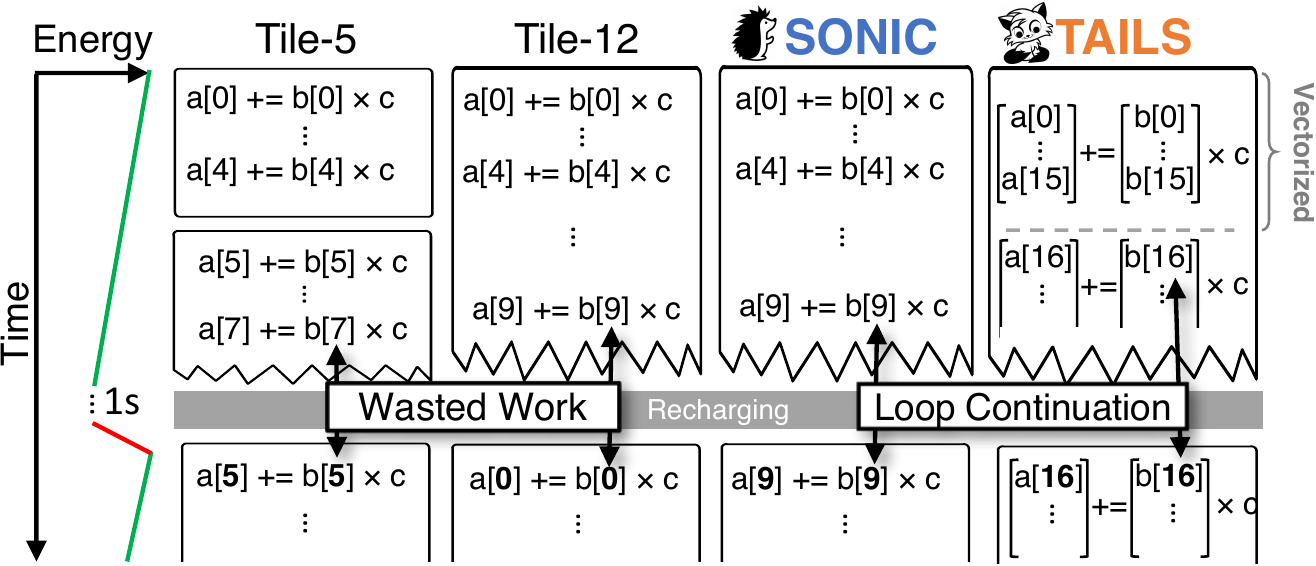}
\caption{\label{fig:looptrick} Executing a loop using two fixed task-tilings and with 
\sonic's loop continuation mechanism. Loop continuation avoids the re-execution and 
non-termination costs of task-tiling.
\tails uses SIMD to perform more work in a fixed energy budget (\autoref{sec:tails}).} 
\vspace{-1em}
\end{figure}

{\em Task-tiling} is a simple way to split a loop's iterations into tasks.  A
task-tiled loop executes a fixed number of iterations per task.  Task-tiling
amortizes task transitioning overhead, but risks executing more iterations in a
single task than the device's energy buffer can support, causing
non-termination.  Figure~\ref{fig:looptrick} shows the intermittent execution
(energy trace on left) of a loop computing a dot product using two
fixed tile sizes of five (Tile-5) and twelve (Tile-12). Tile-5 wastes work when four
iterations complete before a failure.  Tile-12 prevents forward
progress because the device buffers insufficient energy to complete twelve
iterations.

\subsubsection{Loop continuation}

\sonic's \emph{loop continuation} is an intermittence-safe optimization that avoids wasted work,
unnecessary data privatization, and task transition overheads in tasks
containing long-running loop nests. Loop continuation works by directly
modifying loop control variables and memory manipulated in a loop nest, rather
than splitting a long-running loop across tasks. 
Loop continuation
permits loops of arbitrary iteration count within a single task, with neither
non-termination nor excessive state management overhead.
Loop continuation stores a loop's control variables and data manipulated directly in non-volatile
memory \emph{without backing either up}.  When a loop continuation task restarts, its
(volatile) local variables are reinitialized at the task's start.  The loop
control variables, however, retain their state and the loop continues from the
last attempted iteration.

\begin{figure*}[t]
\begin{minipage}{0.49\textwidth}
\centering
\includegraphics[width=\linewidth]{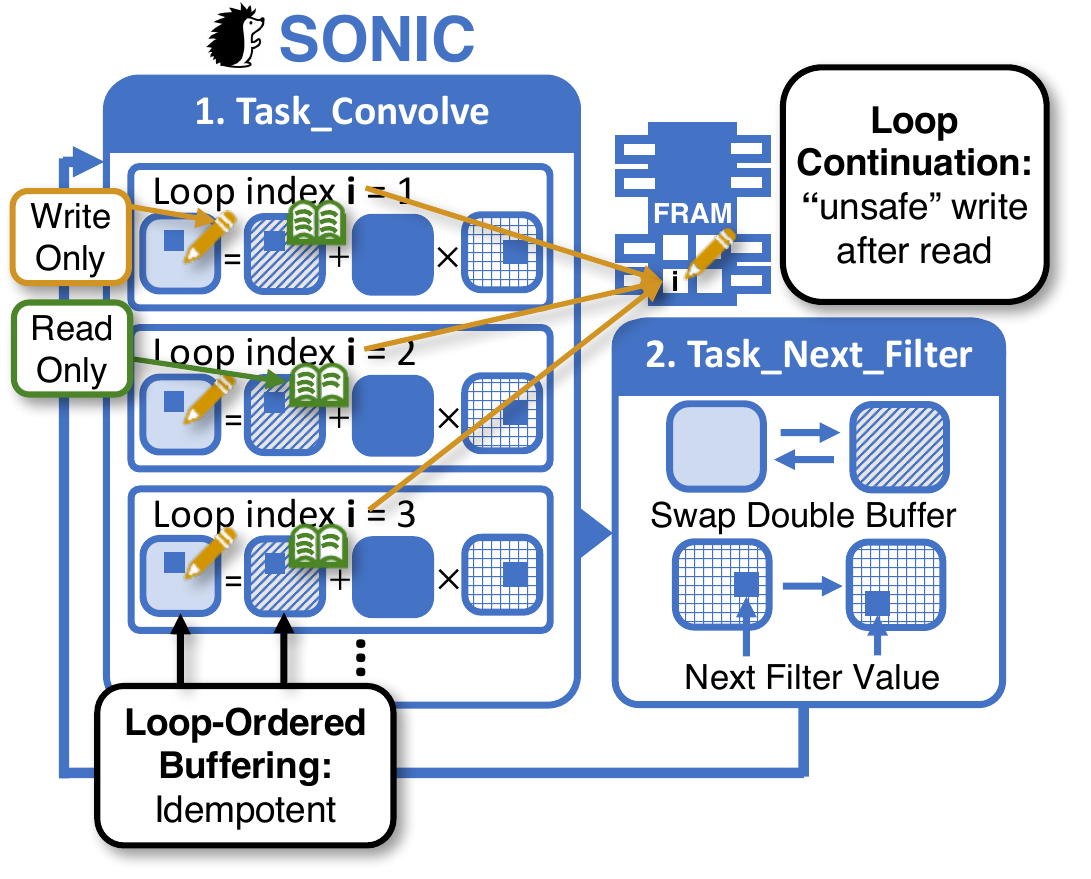}
\caption{\label{fig:tails_sonic}
  \sonic uses \emph{loop continuation} and \emph{loop-ordered buffering} 
  to reduce overheads of correct intermittent execution. \emph{Loop continuation}
  maximizes the amount of computation done per task by allowing computation to
  pick up where it left off before power failure.
  }
  \vspace{-1em}
\end{minipage}
\hfill
\begin{minipage}{0.49\textwidth}
\begin{lstlisting}[caption={\label{listing:sonic_code}Pseudocode corresponding to \autoref{fig:tails_sonic}. 
All variables (except \lstinline{f}) are non-volatile. \lstinline{Task_Convolve} implements loop continuation. 
\lstinline{Task_Next_Filter} atomically swaps buffers and updates variables to move to the next element of the convolutional filter.},captionpos=b]
def Task_Convolve():
	for i in i...len(src): 	# i is NOT reset to 0 here
		f = src[i] * filter[pos]
		dest[i] = (inter[i] + f) if pos > 0 else f

	if pos < len(filter):
		transition Task_Next_Filter
	else:
		pos, i = 0, 0 # reset for next invocation
		transition caller # return from SONIC

def Task_Next_Filter():
	atomic { # swap buffers; reset i; next filter element
		dest, inter = inter, dest
		i = 0
		pos++
	}
	transition Task_Convolve
\end{lstlisting}
\end{minipage}
\end{figure*}

\autoref{fig:tails_sonic} and Listing~\autoref{listing:sonic_code} show how loop continuation
works by storing the loop control state, \textbf{i}, for \lstinline{Task_Convolve} in non-volatile memory.  
\sonic ensures that the loop's control variables \textbf{i} is correct by updating it at the end of the iteration and \emph{not resetting it upon re-execution} (the loop in Listing~\autoref{listing:sonic_code} starts from i).
A power failure during or after the update to
the control variable may require the body of the loop nest to repeat a single
iteration, but %
it never skips an iteration.

Figure~\ref{fig:looptrick} shows \sonic executing using loop
continuation.  Despite the power interruption, execution resumes on the ninth
loop iteration, rather than restarting the entire loop nest or every fifth iteration
like Tile-5 does.

\subsubsection{Idempotence tricks}
\label{sec:sonic:idempotence}

Normally, restarting from the middle of a loop nest could leave manipulated
data partially updated and possibly inconsistent.  However, loop continuation
is safe because \sonic's runtime system ensures each loop iteration is idempotent
using either {\em loop-ordered buffering} or {\em sparse undo-logging}.
\sonic never requires an operation in an iteration to read a value
produced by another operation in the same iteration.
Thus, an iteration that repeatedly re-executes due to power interruption will always
see correct values.

\paragraph{Loop-ordered buffering}
Loop-ordered buffering is a double-buffering mechanism used in convolutional
layers (and dense fully-connected layers)
that ensures each loop iteration is idempotent without expensive redo-logging (cf., \cite{alpaca}). 
Since the MSP430 devices do not possess sophisticated caching
mechanisms, \emph{rather than optimizing for reuse and data locality, \sonic
  optimizes the number of items needed to commit.}
By re-ordering the loops in DNN inference and double-buffering partial activations as needed,
\sonic is able to \emph{completely eliminate} commits within a loop iteration.

Evaluating a sparse or dense convolution requires \sonic to apply a filter to a
layer's entire input activation matrix.  
\sonic orders loop iterations to apply each element of the filter to each
element of the input activation (i.e., multiplying them) before moving on
to the next element of the filter.
For idempotence,
\sonic writes the partially accumulated value to an intermediate output buffer,
rather than applying updates to the input matrix in-place.
After applying a single filter element to each entry in the input and storing
the partial result in the intermediate buffer, \sonic swaps the input
buffer with the intermediate buffer and moves on to the next filter value.

Since \sonic never reads and then writes to the same memory locations within an iteration,
it avoids the WAR problem described in \autoref{sec:background}
and loop iterations are thus idempotent.
\autoref{fig:tails_sonic} shows how under loop-ordered buffering, \sonic never 
reads and writes to the same matrix buffer while computing a partial result in \lstinline{Task_Convolve}
(\lstinline{dest} is distinct from \lstinline{inter} in Listing~\autoref{listing:sonic_code}).
After finishing this task, \sonic transitions to \lstinline{Task_Next_Filter}, which swaps the buffer pointers
and gets the next value to apply from the filter.

\paragraph{Sparse undo-logging}
While loop-ordered buffering is sufficient to ensure each loop iteration is idempotent,
it is sometimes unnecessarily wasteful.
The problem arises because loop-ordered buffering swaps between buffers after every task,
so it must copy data between buffers in case it is read in the future---even if the data has not been modified.
This copying is wasteful on sparse fully-connected layers,
where most filter weights are pruned and thus few activations are modified in a single iteration.
With loop-ordered buffering, \sonic ends up spending most of its time 
and energy copying unmodified activations between buffers.

To eliminate this inefficiency, \sonic introduces \emph{sparse undo-logging}
which ensures idempotence through undo-loggi\-ng instead of double buffering.
To ensure atomicity, sparse undo-logging tracks its progress through the loop via two index variables,
the \emph{read} and \emph{write} indices.
When applying a filter,
\sonic first copies the original, unmodified activation into a canonical memory location,
and then increments the read index.
\sonic then computes the modified activation and writes it back to the original activation buffer
(there is no separate output buffer).
Then it increments the write index and proceeds to the next iteration.
This two-phase approach guarantees correct execution,
since sparse undo-logging resumes computing the output value from the buffered original value if power fails in the middle of an update.

Sparse undo-logging ensures that the work per task grows with the number of modifications made,
not the size of the output buffer (unlike loop-ordered buffering).
However, sparse undo-logging doubles the number of memory writes per modified element,
so it is inefficient on dense layers where most data are modified.
In those cases, loop-ordered buffering is significantly more efficient.
We therefore only use sparse undo-logging in sparse fully-connected layers.
Finally, unlike prior task-based systems such as Alpaca,
sparse undo-logging ensures idempotence with \emph{constant} space overhead
and \emph{no} task transition between iterations.

\paragraph{Related work}
Prior work in persistent memory~\cite{elnawawy2017efficient} uses techniques similar to
our sparse undo-logging.
This work is in the high-performance domain,
and therefore focuses on cache locality and scheduling cache flushes and barriers.
In contrast, our prototype has no caches,
and we exploit this fact in loop-ordered buffering to re-arrange loops in a way that would destroy cache performance on conventional systems.
Moreover, \sonic is more selective than~\cite{elnawawy2017efficient},
only using undo-logging in sparse fully-connected layers where it outperforms double buffering.
\section{Hardware acceleration with \tails}
\label{sec:tails}

\tails improves on \sonic by incorporating widely available hardware acceleration
to perform inference even more efficiently.
A programmer may optionally link their \sonic application to the \tails runtime
system, enabling the application to use direct-memory access (DMA) hardware to
optimize block data movement and to execute operation in parallel using a
simple vector accelerator like the TI Low-Energy Accelerator
(LEA)~\cite{lea}.
LEA supports finite-impulse-response discrete-time convolution (FIR
DTC), which directly implements the convolutions needed in DNN inference.

\tails's runtime system enables the effective use of LEA in an intermittent
system by \emph{adaptively binding
hardware parameters at run time to maximize operational throughput without
exceeding the device's energy buffer}.
Our \tails prototype adaptively determines the DMA block size and LEA vector
width based on the number of operations that successfully complete using the
device's fixed energy buffer.
After calibrating these parameters, \tails uses them to configure available
hardware units and execute inference thereafter. 

\subsection{Automatic one-time calibration}
Before its first execution, a \tails application runs a
short, recursive calibration routine to determine DMA
block size and LEA vector size.  The routine determines the
maximum vector size that it is possible to DMA into LEA's operating buffer,
process using FIR DTC, and DMA back to non-volatile memory without
exceeding the device's energy buffer and impeding progress.  
If a tile size does not complete before power fails, the calibration task re-executes, 
halving the tile size. 
Calibration ends when a FIR DTC completes and \tails uses that tile size for subsequent computations.

\subsection{Accelerating inference with LEA}
Once \tails determines its tile size, the application runs, using DMA and LEA
to compute dense and sparse convolutions and dense matrix multiplications.  
LEA has limitations: it only supports dense operations
and can only read from the device's small 4KB SRAM (not the 256KB FRAM).
\tails uses DMA to move inputs into SRAM,
invokes LEA,
and DMAs the results back to FRAM.
Dense layers are natively supported:
fully-connected layers use LEA's vector MAC operation,
and convolutions use LEA's one-dimensional FIR DTC operation.
To support two- and three-dimensional convolutions, \tails iteratively 
applies one-dimensional convolutions and accumulates those convolutions' results. 
\tails uses loop-ordered buffering to ensure that updates to
the partially accumulated values are idempotent (\autoref{sec:sonic:idempotence}).

Sparse operations require more effort.
\tails uses LEA for sparse convolutions by first making filters dense (padding with zeros).
Making the filters dense is inexpensive %
because each filter is reused many times, amortizing its creation cost.
However, %
this does mean that LEA performs unnecessary work, which sometimes hurts performance.
For this reason, we use LEA's dot-product operation instead of FIR-DTC for $1\times p\times 1$ factored convolutional layers.

Finally, sparse fully-connected layers are inefficient on LEA
because filters do not get reuse.
We found that \tails spent most of its time on padding filters,
and, despite significant effort, we were unable to accelerate sparse fully-connected layers with LEA.
For this reason, \tails performs sparse fully-connected layers in software
exactly like \sonic.

\begin{figure}[h]
\centering
\includegraphics[width=0.9\linewidth]{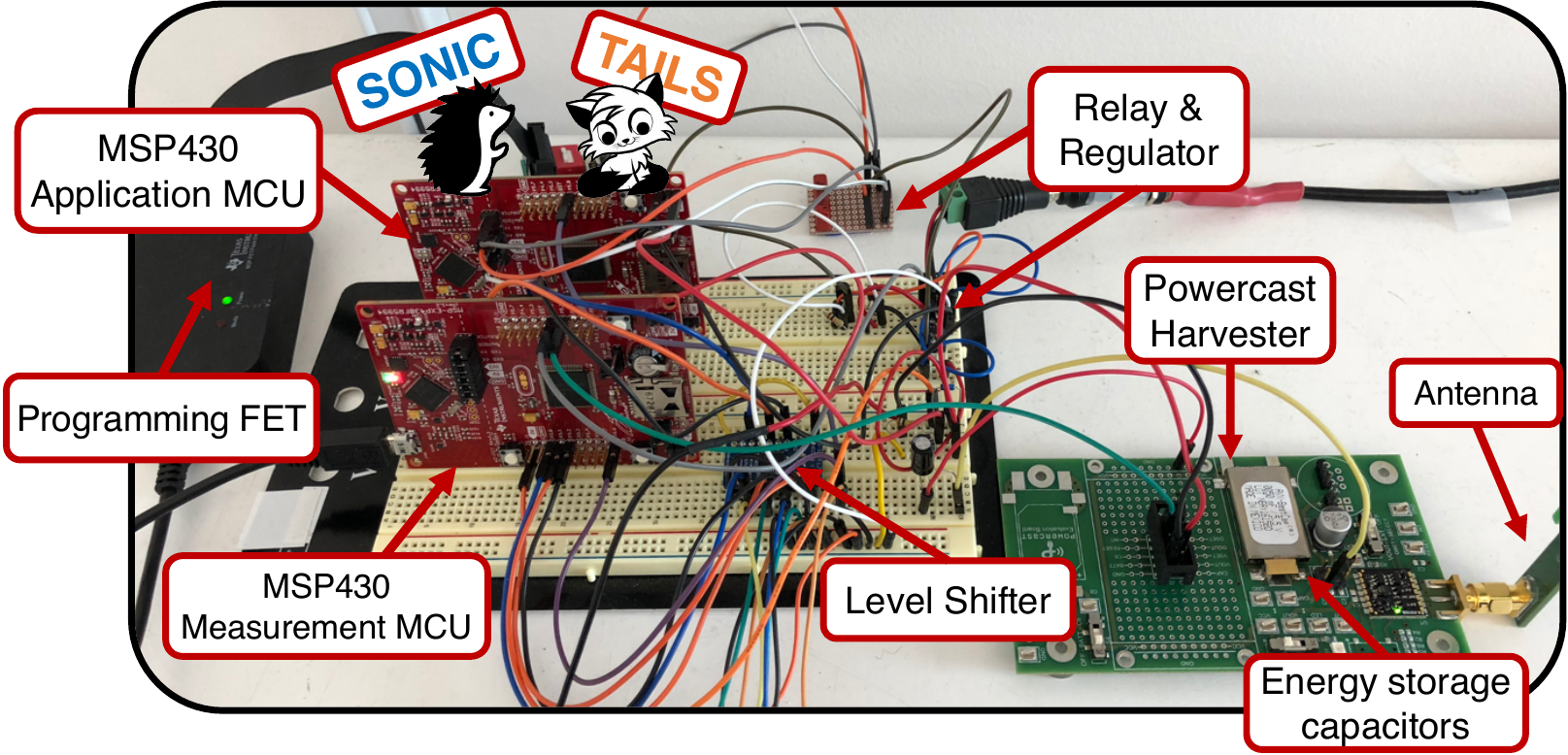}
\caption{\label{fig:hardware} Diagram of the measurement setup.} 
\end{figure}

\figEvalTime
\section{Methodology}
\label{sec:methodology}

We implement \sonic and \tails on the TI-MSP430FR5994~\cite{msp430fr5994} at
16MHz in the setup in \autoref{fig:hardware}.
The board is connected to a
Powercast P2210B~\cite{powercastboard} harvester 1m away from a 3W Powercaster
transmitter~\cite{powercasttransmitter}. We ran all configurations on
continuous power and on intermittent power with three different capacitor
sizes: 1mF, 50mF, and 100\textmu F.

\paragraph{Running code on the device}
We compile with MSPGCC 6.4 and use TI's MSPDriverlib for DMA and TI's DSPLib
for LEA. We use GCC instead of Alpaca's LLVM backend because LLVM lacks support
for 20-bit addressing and produces slower code for MSP430 than GCC.

\paragraph{Measurement}
We use a second MSP430FR5994 to measure intermittent executions.
GPIO pins on the measurement MCU connect through a level-shifter to the
intermittent device, allowing it to count reboots and signal when to start and
stop timing.

We automate measurement with a combination of software and hardware that
compiles a configuration binary, flashes the binary to the device, and
communicates with the measurement MCU to collect results.
The system actuates a relay to switch between continuous power for reprogramming
and intermittent power for testing.

\paragraph{Measuring energy}
By counting the number of charge cycles between GPIO pulses, we can determine
the amount of energy consumed in different code regions.
For a more fine-grained approach, we built a suite of 
microbenchmarks to count how many times a particular operation (e.g., a load from FRAM) can run in single charge cycle.
We then profile how many times each operation is invoked during inference
and scale by per-operation energy to get a detailed energy breakdown.

\paragraph{Baselines for comparison}
We compare \syslong to four DNN inference implementations. The first
implementation is a standard, baseline implementation that does not tolerate
intermittent operation (it does not terminate).
The other three implementations are based on
Alpaca~\cite{alpaca} and split up loops by tiling iterations, as in \autoref{fig:looptrick}.
\section{Evaluation}
\label{sec:evaluation}

We ~now ~evaluate ~our ~prototype ~to ~demonstrate ~that:
\emph{(i)}~\syslong guarantee correct intermittent execution; %
\emph{(ii)}~\syslong greatly reduce inference energy and time over
the state-of-the-art; and
\emph{(iii)}~\syslong perform well across a variety of networks
without any hand-tuning.

\subsection{\syslong significantly accelerate intermittent DNN inference 
over the state-of-the-art}

\autoref{fig:evaluation:time} shows the inference time for the three
networks we consider (\autoref{tab:nns}).
For each network, we evaluated six implementations running on four different power systems.
We break inference time into:
dead time spent recharging;
live time spent on each convolution layer (which dominates);
live time spent on the fully-connected layers;
and everything else.

First, notice that \syslong guarantees correct execution for every
network on every power system.
This is not true of the na\"ive baseline, which does not run correctly
on intermittent power, or of most tilings for prior task-based
intermittent systems.
The only other implementation that reliably executes correctly is
Tile-8, since its tiling is small enough to always complete within a
single charge cycle.
The other tilings fail on some configurations: Tile-32 fails on
MNIST with a 100\textmu F capacitor, and Tile-128 fails on all networks at 100\textmu
F.

\syslong guarantee correct execution at much lower
overheads than Tile-8.
Averaging across networks, Tile-8 is gmean $13.4\times$ slower than the
na\"ive baseline on continuous power, whereas \sonic is $1.45\times$ slower and 
\tails is actually $1.2\times$ \emph{faster} than the baseline.
That is to say, \sonic improves performance on average by $6.9\times$ over tiled Alpaca~\cite{alpaca},
and \tails improves it by $12.2\times$.
Moreover, execution time is consistent across capacitor sizes for \syslong.

\figEvalOther

Larger tile sizes amortize overheads somewhat, but since they do not
complete on all networks or capacitor sizes, they are an unattractive
implementation choice.
\syslong guarantee correct intermittent execution across all capacitor
sizes, while also being faster than the largest tilings: even compared
to Tile-128, \sonic is on average $5.2\times$ faster on continuous power and \tails
is $9.2\times$ faster.

Both DMA and LEA improve \tails's efficiency. We tested configurations 
where DMA and LEA are emulated by software and found that LEA consistently 
improved performance by $1.4\times$, while DMA improved it by $14\%$ on average.

Ultimately, these results indicate that inference is viable on
commodity energy-harvesting devices, and \syslong significantly reduce overheads over
the state-of-the-art. 

\subsection{Loop continuation nearly eliminates overheads due to intermittence}
\autoref{fig:evaluation:time:breakdown} shows that the overheads of \syslong come mainly
from control required to support intermittence. 
The darker-hatched regions of the bars represent the proportion of time spent 
computing a layer's kernel (i.e., the main loop), while the lighter regions represent control overheads (i.e., task transitions and setup/teardown).
Most of the difference in performance between the baseline and \sonic is attributable 
to the lighter, control regions. 
This suggests that \sonic imposes small overhead
over the na\"ive baseline, which accumulates values in registers and avoids memory writes (but does not tolerate intermittence).

\tails's overhead also comes from control; \tails significantly accelerates kernels.
\tails's control overhead is large due to LEA's fixed-point representation,
which forces \tails to bit-shift activations before invoking FIR-DTC.
Moreover, LEA does not have a left-shift operation (it does have a right-shift),
so these shifts must be done in software.
These shifts account for most of the control time in \autoref{fig:evaluation:time:breakdown}.

\autoref{fig:evaluation:time:breakdown} also shows the time breakdown for Tile-32.
Unlike \syslong,
Tile-32 spends significantly more time in both control and the kernel.
This is because Alpaca uses redo-logging on all written values to ensure idempotence,
so every write requires dynamic buffering (kernel time)
and committing when the task completes (control time).
\syslong effectively eliminate redo-logging, avoiding these overheads.

\subsection{\syslong use much less energy than tiling}

Energy-harvesting systems spend a majority of their time powered off
recharging, so execution time is largely determined by energy
efficiency.
\autoref{fig:evaluation:energy:measured} shows that \syslong achieve high performance 
because they require less energy than other schemes.
Inference energy is in direct proportion to the dead
time spent recharging in \autoref{fig:evaluation:time}.
Since dead time dominates inference time, \syslong get similar
improvements in inference energy as they do in terms of inference
time.

\subsection{Where does \sonic's energy go?}
\autoref{fig:evaluation:energy:micro} further characterizes \sonic by showing 
the proportion of energy spent on different operations. 
The blue regions represent memory operations, the orange regions are control
instructions, the green regions are arithmetic instructions within the kernels,
the purple regions are the task-transition overhead, and the grey regions
are the remaining, unaccounted-for energy. 
The control instructions account for 26\% of \sonic's energy,
and a further 14\% of system energy comes from FRAM writes to loop indices.
Ideally, these overheads would be amortized across many kernel operations,
but doing this requires a more efficient architecture.

\vspace{-0.5em}
\section{Future intermittent architecture research}
\label{sec:panic}

Our experience in building \syslong demonstrates there is a large
opportunity to accelerate intermittent inference via
a parallel architecture with built-in support for intermittent operation.
However, typical microcontrollers for energy-harvesting systems are poorly
suited to efficient inference, and we have identified several opportunities to
significantly improve them with better hardware support.
Current microcontrollers are sequential, single-cycle processors,
and so spend very little of their energy on ``useful work''~\cite{horowitz:isscc14:energy-keynote}.
For example, by deducting the energy of \texttt{nop} instructions from \autoref{fig:evaluation:energy:micro},
we estimate that \sonic spends 40\% of its energy on instruction fetch and decode.
This cost is a waste in highly structured computations like DNN inference,
where overheads easily amortize over many operations. %

LEA should bridge this efficiency gap, but unfortunately LEA has many limitations.
Invoking LEA is expensive. Each LEA invocation should therefore do as much work as
possible, but LEA's parallelism is limited by its small (4KB) SRAM buffer.
This small buffer also forces frequent DMA between SRAM and FRAM,
which cannot be overlapped with LEA execution
and does not support strided accesses or scatter-gather.
LEA has surprising gaps in its support:
it does not support vector left-shift or scalar multiply,
forcing \tails to fall back to software.
In software, integer multiplication is a memory-mapped peripheral that takes
four instructions and nine cycles.
All told, these limitations cause \syslong to spend much more energy than necessary.
There is ample room to improve inference efficiency via a better architecture.

Thus far, architectures for intermittent computing
have focused on how hardware can efficiently guarantee correctness~\cite{clank,nvp,ma2017incidental}.
While there is certainly scope for architectural support,
correctness requires a full-stack approach.
Handling correctness in the architecture alone is insufficient because it ignores
system-level effects, such as I/O (e.g., sensors and radios),
data timeliness, and interrupts that must be part of an
end-to-end correctness guarantee~\cite{mayfly,capybara}.
Moreover, an architecture-only approach is energy-inefficient because it must
conservatively back up architectural state in non-volatile memory after each instruction.
Software can instead identify precisely what state is needed for correctness (e.g., loop
indices in \sonic).
We therefore see more opportunity in targeted architectural support
(e.g., caches with just-in-time checkpointing to avoid frequent, expensive writes to non-volatile memory for index variables),
than in conservative models that ask nothing of software~\cite{nvp,ma2017incidental}.

Furthermore, to enable compute-heavy applications like inference and signal processing,
future intermittent architectures must aggressively optimize for energy efficiency.
The key is to eliminate or amortize wasted energy %
(e.g., in fetch, decode, register file, and FRAM)---%
we estimate that a new architecture would save 14\% of system energy
just by eliminating frequent FRAM writes to loop indices alone!

Intermittent architectures must navigate several fundamental design challenges
to optimize energy efficiency.
Highly specialized architectures (e.g., ASICs) are the most efficient,
but sacrifice programmability.
Such specialization is premature in intermittent computing systems
because the dominant applications in this domain are yet to be determined;
programmability remains essential.
Programmable architectures can achieve ASIC-like efficiency on highly parallel codes
by amortizing energy spent across many in-flight operations.
Unfortunately, this requires high power and large amounts of state~\cite{culler:isca88:parallelism-resources},
both of which are non-starters in energy-harvesting systems.
Hence, a balance of modest specialization and SIMD parallelism is needed
to maximize energy-efficiency~\cite{hameed2010understanding,dally:ieee08:elm}.
We are currently exploring an intermittent parallel architecture inspired by streaming dataflow models~\cite{nowatzki2015exploring,dyser,plasticine,trips},
striking an appealing balance between programmability, parallelism, and specialization
to maximize efficiency without compromising the architecture's ability to provide correctness guarantees.

\section{Conclusion}
\label{sec:conclusion}

This paper has argued that intelligence ``beyond the edge'' will enable
new classes of IoT applications, and presented the first
demonstration of efficient DNN inference on commodity energy-harvesting systems.
We presented a high-level analysis of why inference accuracy matters,
and used this analysis to automatically compress
networks to maximize end-to-end application performance.
\syslong then specialize intermittence support to guarantee correct
execution, regardless of power system, while reducing overheads by
up to $6.9\times$ and $12.2\times$, respectively, over the state-of-the-art.
\section{Acknowledgements}
\label{sec:acknowledgements}

This work was generously funded by National Science Foundation Awards CCF-1815882 and CCF-1751029.
 
\clearpage
\flushend

\bibliographystyle{ACM-Reference-Format}
\bibliography{refs,confs}


\begin{thebibliography}{81}


\ifx \showCODEN    \undefined \def \showCODEN     #1{\unskip}     \fi
\ifx \showDOI      \undefined \def \showDOI       #1{#1}\fi
\ifx \showISBNx    \undefined \def \showISBNx     #1{\unskip}     \fi
\ifx \showISBNxiii \undefined \def \showISBNxiii  #1{\unskip}     \fi
\ifx \showISSN     \undefined \def \showISSN      #1{\unskip}     \fi
\ifx \showLCCN     \undefined \def \showLCCN      #1{\unskip}     \fi
\ifx \shownote     \undefined \def \shownote      #1{#1}          \fi
\ifx \showarticletitle \undefined \def \showarticletitle #1{#1}   \fi
\ifx \showURL      \undefined \def \showURL       {\relax}        \fi
\providecommand\bibfield[2]{#2}
\providecommand\bibinfo[2]{#2}
\providecommand\natexlab[1]{#1}
\providecommand\showeprint[2][]{arXiv:#2}

\bibitem[\protect\citeauthoryear{??}{lea}{[n. d.]}]%
        {lea}
 \bibinfo{year}{[n. d.]}\natexlab{}.
\newblock \bibinfo{title}{Low Energy Accelerator FAQ}.
\newblock
\newblock
\urldef\tempurl%
\url{http://www.ti.com/lit/an/slaa720/slaa720.pdf}
\showURL{%
\tempurl}


\bibitem[\protect\citeauthoryear{??}{msp}{[n. d.]}]%
        {msp430fr5994}
 \bibinfo{year}{[n. d.]}\natexlab{}.
\newblock \bibinfo{title}{MSP430fr5994 SLA}.
\newblock
\newblock
\urldef\tempurl%
\url{http://www.ti.com/lit/ds/symlink/msp430fr5994.pdf}
\showURL{%
\tempurl}


\bibitem[\protect\citeauthoryear{??}{jet}{[n. d.]}]%
        {jetsontx2}
 \bibinfo{year}{[n. d.]}\natexlab{}.
\newblock \bibinfo{title}{Nividia Jetson TX2}.
\newblock
\newblock
\urldef\tempurl%
\url{https://developer.nvidia.com/embedded/develop/hardware}
\showURL{%
\tempurl}


\bibitem[\protect\citeauthoryear{??}{pow}{[n. d.]a}]%
        {powercastboard}
 \bibinfo{year}{[n. d.]}\natexlab{a}.
\newblock \bibinfo{title}{Powercast P2110B}.
\newblock
\newblock
\urldef\tempurl%
\url{http://www.powercastco.com/wp-content/uploads/2016/12/P2110B-Datasheet-Rev-3.pdf}
\showURL{%
\tempurl}


\bibitem[\protect\citeauthoryear{??}{pow}{[n. d.]b}]%
        {powercasttransmitter}
 \bibinfo{year}{[n. d.]}\natexlab{b}.
\newblock \bibinfo{title}{Powercaster Transmitter}.
\newblock
\newblock
\urldef\tempurl%
\url{http://www.powercastco.com/wp-content/uploads/2016/11/User-Manual-TX-915-01-Rev-A-4.pdf}
\showURL{%
\tempurl}


\bibitem[\protect\citeauthoryear{Albericio, Judd, Hetherington, Aamodt, Jerger,
  and Moshovos}{Albericio et~al\mbox{.}}{2016}]%
        {albericio2016cnvlutin}
\bibfield{author}{\bibinfo{person}{Jorge Albericio}, \bibinfo{person}{Patrick
  Judd}, \bibinfo{person}{Tayler Hetherington}, \bibinfo{person}{Tor Aamodt},
  \bibinfo{person}{Natalie~Enright Jerger}, {and} \bibinfo{person}{Andreas
  Moshovos}.} \bibinfo{year}{2016}\natexlab{}.
\newblock \showarticletitle{Cnvlutin: Ineffectual-neuron-free deep neural
  network computing}. In \bibinfo{booktitle}{\emph{ACM SIGARCH Computer
  Architecture News}}, Vol.~\bibinfo{volume}{44}. IEEE Press,
  \bibinfo{pages}{1--13}.
\newblock


\bibitem[\protect\citeauthoryear{Alwani, Chen, Ferdman, and Milder}{Alwani
  et~al\mbox{.}}{2016}]%
        {alwani2016fused}
\bibfield{author}{\bibinfo{person}{Manoj Alwani}, \bibinfo{person}{Han Chen},
  \bibinfo{person}{Michael Ferdman}, {and} \bibinfo{person}{Peter Milder}.}
  \bibinfo{year}{2016}\natexlab{}.
\newblock \showarticletitle{Fused-layer CNN accelerators}. In
  \bibinfo{booktitle}{\emph{Microarchitecture (MICRO), 2016 49th Annual
  IEEE/ACM International Symposium on}}. IEEE, \bibinfo{pages}{1--12}.
\newblock


\bibitem[\protect\citeauthoryear{{Angus Galloway}}{{Angus Galloway}}{2018}]%
        {binarynetgithub}
\bibfield{author}{\bibinfo{person}{{Angus Galloway}}.}
  \bibinfo{year}{2018}\natexlab{}.
\newblock \bibinfo{title}{{Tensorflow XNOR-BNN}}.
\newblock
  \bibinfo{howpublished}{\url{https://github.com/AngusG/tensorflow-xnor-bnn}}.
\newblock


\bibitem[\protect\citeauthoryear{Bhattacharya and Lane}{Bhattacharya and
  Lane}{2016}]%
        {bhattacharya2016sparsification}
\bibfield{author}{\bibinfo{person}{Sourav Bhattacharya} {and}
  \bibinfo{person}{Nicholas~D Lane}.} \bibinfo{year}{2016}\natexlab{}.
\newblock \showarticletitle{Sparsification and separation of deep learning
  layers for constrained resource inference on wearables}. In
  \bibinfo{booktitle}{\emph{Proceedings of the 14th ACM Conference on Embedded
  Network Sensor Systems CD-ROM}}. ACM, \bibinfo{pages}{176--189}.
\newblock


\bibitem[\protect\citeauthoryear{Buettner, Greenstein, and Wetherall}{Buettner
  et~al\mbox{.}}{2011}]%
        {dewdrop}
\bibfield{author}{\bibinfo{person}{Michael Buettner}, \bibinfo{person}{Ben
  Greenstein}, {and} \bibinfo{person}{David Wetherall}.}
  \bibinfo{year}{2011}\natexlab{}.
\newblock \showarticletitle{Dewdrop: An Energy-Aware Task Scheduler for
  Computational {RFID}}. In \bibinfo{booktitle}{\emph{USENIX Symposium on
  Networked Systems Design and Implementation (NSDI)}}.
\newblock


\bibitem[\protect\citeauthoryear{Chen, Du, Sun, Wang, Wu, Chen, and Temam}{Chen
  et~al\mbox{.}}{2014a}]%
        {chen:asplos14:diannao}
\bibfield{author}{\bibinfo{person}{Tianshi Chen}, \bibinfo{person}{Zidong Du},
  \bibinfo{person}{Ninghui Sun}, \bibinfo{person}{Jia Wang},
  \bibinfo{person}{Chengyong Wu}, \bibinfo{person}{Yunji Chen}, {and}
  \bibinfo{person}{Olivier Temam}.} \bibinfo{year}{2014}\natexlab{a}.
\newblock \showarticletitle{{DianNao}: a small-footprint high-throughput
  accelerator for ubiquitous machine-learning}. In
  \bibinfo{booktitle}{\emph{Proc. of the 19th intl. conf. on Architectural
  Support for Programming Languages and Operating Systems}}.
\newblock


\bibitem[\protect\citeauthoryear{Chen, Luo, Liu, Zhang, He, Wang, Li, Chen, Xu,
  Sun, et~al\mbox{.}}{Chen et~al\mbox{.}}{2014b}]%
        {chen2014dadiannao}
\bibfield{author}{\bibinfo{person}{Yunji Chen}, \bibinfo{person}{Tao Luo},
  \bibinfo{person}{Shaoli Liu}, \bibinfo{person}{Shijin Zhang},
  \bibinfo{person}{Liqiang He}, \bibinfo{person}{Jia Wang},
  \bibinfo{person}{Ling Li}, \bibinfo{person}{Tianshi Chen},
  \bibinfo{person}{Zhiwei Xu}, \bibinfo{person}{Ninghui Sun}, {et~al\mbox{.}}}
  \bibinfo{year}{2014}\natexlab{b}.
\newblock \showarticletitle{Dadiannao: A machine-learning supercomputer}. In
  \bibinfo{booktitle}{\emph{Proceedings of the 47th Annual IEEE/ACM
  International Symposium on Microarchitecture}}. IEEE Computer Society,
  \bibinfo{pages}{609--622}.
\newblock


\bibitem[\protect\citeauthoryear{Chen, Emer, and Sze}{Chen
  et~al\mbox{.}}{2016}]%
        {chen:isca16:eyeriss}
\bibfield{author}{\bibinfo{person}{Yu-Hsin Chen}, \bibinfo{person}{Joel Emer},
  {and} \bibinfo{person}{Vivienne Sze}.} \bibinfo{year}{2016}\natexlab{}.
\newblock \showarticletitle{Eyeriss: A spatial architecture for
  energy-efficient dataflow for convolutional neural networks}. In
  \bibinfo{booktitle}{\emph{Proc. of the 43rd annual Intl. Symp. on Computer
  Architecture (Proc. ISCA-43)}}.
\newblock


\bibitem[\protect\citeauthoryear{Chollet}{Chollet}{[n. d.]}]%
        {chollet2016xception}
\bibfield{author}{\bibinfo{person}{Fran{\c{c}}ois Chollet}.} \bibinfo{year}{[n.
  d.]}\natexlab{}.
\newblock \showarticletitle{Xception: Deep learning with depthwise separable
  convolutions}.
\newblock  (\bibinfo{year}{[n. d.]}).
\newblock


\bibitem[\protect\citeauthoryear{Colin, Harvey, Lucia, and Sample}{Colin
  et~al\mbox{.}}{2016}]%
        {edb}
\bibfield{author}{\bibinfo{person}{Alexei Colin}, \bibinfo{person}{Graham
  Harvey}, \bibinfo{person}{Brandon Lucia}, {and} \bibinfo{person}{Alanson~P.
  Sample}.} \bibinfo{year}{2016}\natexlab{}.
\newblock \showarticletitle{An Energy-interference-free Hardware-Software
  Debugger for Intermittent Energy-harvesting Systems}.
\newblock \bibinfo{journal}{\emph{SIGOPS Oper. Syst. Rev.}}
  \bibinfo{volume}{50}, \bibinfo{number}{2} (\bibinfo{date}{March}
  \bibinfo{year}{2016}), \bibinfo{pages}{577--589}.
\newblock
\showISSN{0163-5980}
\urldef\tempurl%
\url{https://doi.org/10.1145/2954680.2872409}
\showDOI{\tempurl}


\bibitem[\protect\citeauthoryear{Colin and Lucia}{Colin and Lucia}{2016}]%
        {chain}
\bibfield{author}{\bibinfo{person}{Alexei Colin} {and} \bibinfo{person}{Brandon
  Lucia}.} \bibinfo{year}{2016}\natexlab{}.
\newblock \showarticletitle{Chain: Tasks and Channels for Reliable Intermittent
  Programs}. In \bibinfo{booktitle}{\emph{Proceedings of the ACM International
  Conference on Object Oriented Programming Systems Languages and Applications
  (OOPSLA)}}.
\newblock


\bibitem[\protect\citeauthoryear{Colin, Ruppel, and Lucia}{Colin
  et~al\mbox{.}}{2018}]%
        {capybara}
\bibfield{author}{\bibinfo{person}{Alexei Colin}, \bibinfo{person}{Emily
  Ruppel}, {and} \bibinfo{person}{Brandon Lucia}.}
  \bibinfo{year}{2018}\natexlab{}.
\newblock \showarticletitle{A Reconfigurable Energy Storage Architecture for
  Energy-harvesting Devices}. In \bibinfo{booktitle}{\emph{ASPLOS}}.
\newblock


\bibitem[\protect\citeauthoryear{Courbariaux, Hubara, Soudry, El-Yaniv, and
  Bengio}{Courbariaux et~al\mbox{.}}{2016}]%
        {courbariaux2016binarized}
\bibfield{author}{\bibinfo{person}{Matthieu Courbariaux}, \bibinfo{person}{Itay
  Hubara}, \bibinfo{person}{Daniel Soudry}, \bibinfo{person}{Ran El-Yaniv},
  {and} \bibinfo{person}{Yoshua Bengio}.} \bibinfo{year}{2016}\natexlab{}.
\newblock \showarticletitle{Binarized neural networks: Training deep neural
  networks with weights and activations constrained to+ 1 or-1}.
\newblock \bibinfo{journal}{\emph{arXiv preprint arXiv:1602.02830}}
  (\bibinfo{year}{2016}).
\newblock


\bibitem[\protect\citeauthoryear{Culler et~al\mbox{.}}{Culler
  et~al\mbox{.}}{1988}]%
        {culler:isca88:parallelism-resources}
\bibfield{author}{\bibinfo{person}{David~E Culler} {et~al\mbox{.}}}
  \bibinfo{year}{1988}\natexlab{}.
\newblock \showarticletitle{Resource requirements of dataflow programs}. In
  \bibinfo{booktitle}{\emph{ACM SIGARCH Computer Architecture News}},
  Vol.~\bibinfo{volume}{16}. IEEE Computer Society Press,
  \bibinfo{pages}{141--150}.
\newblock


\bibitem[\protect\citeauthoryear{Dally, Balfour, Black-Shaffer, Chen, Harting,
  Parikh, Park, and Sheffield}{Dally et~al\mbox{.}}{2008}]%
        {dally:ieee08:elm}
\bibfield{author}{\bibinfo{person}{William~J Dally}, \bibinfo{person}{James
  Balfour}, \bibinfo{person}{David Black-Shaffer}, \bibinfo{person}{James
  Chen}, \bibinfo{person}{R~Curtis Harting}, \bibinfo{person}{Vishal Parikh},
  \bibinfo{person}{Jongsoo Park}, {and} \bibinfo{person}{David Sheffield}.}
  \bibinfo{year}{2008}\natexlab{}.
\newblock \showarticletitle{Efficient embedded computing}.
\newblock \bibinfo{journal}{\emph{Computer}} \bibinfo{volume}{41},
  \bibinfo{number}{7} (\bibinfo{year}{2008}).
\newblock


\bibitem[\protect\citeauthoryear{De~Lathauwer, De~Moor, and
  Vandewalle}{De~Lathauwer et~al\mbox{.}}{2000a}]%
        {de2000multilinear}
\bibfield{author}{\bibinfo{person}{Lieven De~Lathauwer}, \bibinfo{person}{Bart
  De~Moor}, {and} \bibinfo{person}{Joos Vandewalle}.}
  \bibinfo{year}{2000}\natexlab{a}.
\newblock \showarticletitle{A multilinear singular value decomposition}.
\newblock \bibinfo{journal}{\emph{SIAM journal on Matrix Analysis and
  Applications}} \bibinfo{volume}{21}, \bibinfo{number}{4}
  (\bibinfo{year}{2000}), \bibinfo{pages}{1253--1278}.
\newblock


\bibitem[\protect\citeauthoryear{De~Lathauwer, De~Moor, and
  Vandewalle}{De~Lathauwer et~al\mbox{.}}{2000b}]%
        {de2000best}
\bibfield{author}{\bibinfo{person}{Lieven De~Lathauwer}, \bibinfo{person}{Bart
  De~Moor}, {and} \bibinfo{person}{Joos Vandewalle}.}
  \bibinfo{year}{2000}\natexlab{b}.
\newblock \showarticletitle{On the best rank-1 and rank-(r 1, r 2,..., rn)
  approximation of higher-order tensors}.
\newblock \bibinfo{journal}{\emph{SIAM journal on Matrix Analysis and
  Applications}} \bibinfo{volume}{21}, \bibinfo{number}{4}
  (\bibinfo{year}{2000}), \bibinfo{pages}{1324--1342}.
\newblock


\bibitem[\protect\citeauthoryear{De~Sa, Feldman, R{\'e}, and Olukotun}{De~Sa
  et~al\mbox{.}}{2017}]%
        {desa:isca17:sgd}
\bibfield{author}{\bibinfo{person}{Christopher De~Sa}, \bibinfo{person}{Matthew
  Feldman}, \bibinfo{person}{Christopher R{\'e}}, {and} \bibinfo{person}{Kunle
  Olukotun}.} \bibinfo{year}{2017}\natexlab{}.
\newblock \showarticletitle{Understanding and Optimizing Asynchronous
  Low-Precision Stochastic Gradient Descent}. In
  \bibinfo{booktitle}{\emph{Proc. of the 44th annual Intl. Symp. on Computer
  Architecture (Proc. ISCA-44)}}.
\newblock


\bibitem[\protect\citeauthoryear{Ding, Liao, Wang, Li, Liu, Zhuo, Wang, Qian,
  Bai, Yuan, et~al\mbox{.}}{Ding et~al\mbox{.}}{2017}]%
        {ding2017circnn}
\bibfield{author}{\bibinfo{person}{Caiwen Ding}, \bibinfo{person}{Siyu Liao},
  \bibinfo{person}{Yanzhi Wang}, \bibinfo{person}{Zhe Li},
  \bibinfo{person}{Ning Liu}, \bibinfo{person}{Youwei Zhuo},
  \bibinfo{person}{Chao Wang}, \bibinfo{person}{Xuehai Qian},
  \bibinfo{person}{Yu Bai}, \bibinfo{person}{Geng Yuan}, {et~al\mbox{.}}}
  \bibinfo{year}{2017}\natexlab{}.
\newblock \showarticletitle{CirCNN: accelerating and compressing deep neural
  networks using block-circulant weight matrices}. In
  \bibinfo{booktitle}{\emph{Proceedings of the 50th Annual IEEE/ACM
  International Symposium on Microarchitecture}}. ACM,
  \bibinfo{pages}{395--408}.
\newblock


\bibitem[\protect\citeauthoryear{Dongare, Hesling, Bhatia, Balanuta, Pereira,
  Iannucci, and Rowe}{Dongare et~al\mbox{.}}{2017}]%
        {dongare2017openchirp}
\bibfield{author}{\bibinfo{person}{Adwait Dongare}, \bibinfo{person}{Craig
  Hesling}, \bibinfo{person}{Khushboo Bhatia}, \bibinfo{person}{Artur
  Balanuta}, \bibinfo{person}{Ricardo~Lopes Pereira}, \bibinfo{person}{Bob
  Iannucci}, {and} \bibinfo{person}{Anthony Rowe}.}
  \bibinfo{year}{2017}\natexlab{}.
\newblock \showarticletitle{OpenChirp: A low-power wide-area networking
  architecture}. In \bibinfo{booktitle}{\emph{Pervasive Computing and
  Communications Workshops (PerCom Workshops), 2017 IEEE International
  Conference on}}. IEEE, \bibinfo{pages}{569--574}.
\newblock


\bibitem[\protect\citeauthoryear{Du, Fasthuber, Chen, Ienne, Li, Luo, Feng,
  Chen, and Temam}{Du et~al\mbox{.}}{2015}]%
        {du:isca15:shidiannao}
\bibfield{author}{\bibinfo{person}{Zidong Du}, \bibinfo{person}{Robert
  Fasthuber}, \bibinfo{person}{Tianshi Chen}, \bibinfo{person}{Paolo Ienne},
  \bibinfo{person}{Ling Li}, \bibinfo{person}{Tao Luo},
  \bibinfo{person}{Xiaobing Feng}, \bibinfo{person}{Yunji Chen}, {and}
  \bibinfo{person}{Olivier Temam}.} \bibinfo{year}{2015}\natexlab{}.
\newblock \showarticletitle{ShiDianNao: Shifting vision processing closer to
  the sensor}. In \bibinfo{booktitle}{\emph{Proc. of the 42nd annual Intl.
  Symp. on Computer Architecture (Proc. ISCA-42)}}.
\newblock


\bibitem[\protect\citeauthoryear{Elnawawy, Alshboul, Tuck, and
  Solihin}{Elnawawy et~al\mbox{.}}{2017}]%
        {elnawawy2017efficient}
\bibfield{author}{\bibinfo{person}{Hussein Elnawawy}, \bibinfo{person}{Mohammad
  Alshboul}, \bibinfo{person}{James Tuck}, {and} \bibinfo{person}{Yan
  Solihin}.} \bibinfo{year}{2017}\natexlab{}.
\newblock \showarticletitle{Efficient Checkpointing of Loop-Based Codes for
  Non-volatile Main Memory}. In \bibinfo{booktitle}{\emph{Parallel
  Architectures and Compilation Techniques (PACT), 2017 26th International
  Conference on}}. IEEE, \bibinfo{pages}{318--329}.
\newblock


\bibitem[\protect\citeauthoryear{Fick, Blaauw, Sylvester, Skrzyniarz, Parikh,
  and Fick}{Fick et~al\mbox{.}}{2017}]%
        {fick2017subthresholdinference}
\bibfield{author}{\bibinfo{person}{L. Fick}, \bibinfo{person}{D. Blaauw},
  \bibinfo{person}{D. Sylvester}, \bibinfo{person}{S. Skrzyniarz},
  \bibinfo{person}{M. Parikh}, {and} \bibinfo{person}{D. Fick}.}
  \bibinfo{year}{2017}\natexlab{}.
\newblock \showarticletitle{Analog in-memory subthreshold deep neural network
  accelerator}. In \bibinfo{booktitle}{\emph{2017 IEEE Custom Integrated
  Circuits Conference (CICC)}}. \bibinfo{pages}{1--4}.
\newblock
\urldef\tempurl%
\url{https://doi.org/10.1109/CICC.2017.7993629}
\showDOI{\tempurl}


\bibitem[\protect\citeauthoryear{Gobieski, Beckmann, and Lucia}{Gobieski
  et~al\mbox{.}}{2018}]%
        {Gobieski2018IntermittentDN}
\bibfield{author}{\bibinfo{person}{Graham Gobieski}, \bibinfo{person}{Nathan
  Beckmann}, {and} \bibinfo{person}{Brandon Lucia}.}
  \bibinfo{year}{2018}\natexlab{}.
\newblock \showarticletitle{Intermittent Deep Neural Network Inference}. In
  \bibinfo{booktitle}{\emph{SysML}}.
\newblock


\bibitem[\protect\citeauthoryear{Golovin, Solnik, Moitra, Kochanski, Karro, and
  Sculley}{Golovin et~al\mbox{.}}{2017}]%
        {golovin2017google}
\bibfield{author}{\bibinfo{person}{Daniel Golovin}, \bibinfo{person}{Benjamin
  Solnik}, \bibinfo{person}{Subhodeep Moitra}, \bibinfo{person}{Greg
  Kochanski}, \bibinfo{person}{John Karro}, {and} \bibinfo{person}{D Sculley}.}
  \bibinfo{year}{2017}\natexlab{}.
\newblock \showarticletitle{Google vizier: A service for black-box
  optimization}. In \bibinfo{booktitle}{\emph{Proceedings of the 23rd ACM
  SIGKDD International Conference on Knowledge Discovery and Data Mining}}.
  ACM, \bibinfo{pages}{1487--1495}.
\newblock


\bibitem[\protect\citeauthoryear{Govindaraju, Ho, Nowatzki, Chhugani, Satish,
  Sankaralingam, and Kim}{Govindaraju et~al\mbox{.}}{2012}]%
        {dyser}
\bibfield{author}{\bibinfo{person}{Venkatraman Govindaraju},
  \bibinfo{person}{Chen-Han Ho}, \bibinfo{person}{Tony Nowatzki},
  \bibinfo{person}{Jatin Chhugani}, \bibinfo{person}{Nadathur Satish},
  \bibinfo{person}{Karthikeyan Sankaralingam}, {and} \bibinfo{person}{Changkyu
  Kim}.} \bibinfo{year}{2012}\natexlab{}.
\newblock \showarticletitle{Dyser: Unifying functionality and parallelism
  specialization for energy-efficient computing}.
\newblock \bibinfo{journal}{\emph{IEEE Micro}} \bibinfo{volume}{32},
  \bibinfo{number}{5} (\bibinfo{year}{2012}), \bibinfo{pages}{38--51}.
\newblock


\bibitem[\protect\citeauthoryear{Gupta, Suggala, Goyal, Simhadri, Paranjape,
  Kumar, Goyal, Udupa, Varma, and Jain}{Gupta et~al\mbox{.}}{2017}]%
        {gupta2017protonn}
\bibfield{author}{\bibinfo{person}{Chirag Gupta}, \bibinfo{person}{Arun~Sai
  Suggala}, \bibinfo{person}{Ankit Goyal}, \bibinfo{person}{Harsha~Vardhan
  Simhadri}, \bibinfo{person}{Bhargavi Paranjape}, \bibinfo{person}{Ashish
  Kumar}, \bibinfo{person}{Saurabh Goyal}, \bibinfo{person}{Raghavendra Udupa},
  \bibinfo{person}{Manik Varma}, {and} \bibinfo{person}{Prateek Jain}.}
  \bibinfo{year}{2017}\natexlab{}.
\newblock \showarticletitle{ProtoNN: Compressed and Accurate kNN for
  Resource-scarce Devices}. In \bibinfo{booktitle}{\emph{International
  Conference on Machine Learning}}. \bibinfo{pages}{1331--1340}.
\newblock


\bibitem[\protect\citeauthoryear{Hameed, Qadeer, Wachs, Azizi, Solomatnikov,
  Lee, Richardson, Kozyrakis, and Horowitz}{Hameed et~al\mbox{.}}{2010}]%
        {hameed2010understanding}
\bibfield{author}{\bibinfo{person}{Rehan Hameed}, \bibinfo{person}{Wajahat
  Qadeer}, \bibinfo{person}{Megan Wachs}, \bibinfo{person}{Omid Azizi},
  \bibinfo{person}{Alex Solomatnikov}, \bibinfo{person}{Benjamin~C Lee},
  \bibinfo{person}{Stephen Richardson}, \bibinfo{person}{Christos Kozyrakis},
  {and} \bibinfo{person}{Mark Horowitz}.} \bibinfo{year}{2010}\natexlab{}.
\newblock \showarticletitle{Understanding sources of inefficiency in
  general-purpose chips}. In \bibinfo{booktitle}{\emph{ACM SIGARCH Computer
  Architecture News}}, Vol.~\bibinfo{volume}{38}. ACM, \bibinfo{pages}{37--47}.
\newblock


\bibitem[\protect\citeauthoryear{Han, Liu, Mao, Pu, Pdream, Horowitz, and
  Dally}{Han et~al\mbox{.}}{2016a}]%
        {han:isca16:eie}
\bibfield{author}{\bibinfo{person}{Song Han}, \bibinfo{person}{Xingyu Liu},
  \bibinfo{person}{Huizi Mao}, \bibinfo{person}{Jing Pu},
  \bibinfo{person}{Ardavan Pdream}, \bibinfo{person}{Mark~A. Horowitz}, {and}
  \bibinfo{person}{William~J. Dally}.} \bibinfo{year}{2016}\natexlab{a}.
\newblock \showarticletitle{EIE: Efficient Inference Engine on Compressed Deep
  Neural Network}. In \bibinfo{booktitle}{\emph{Proc. of the 43rd annual Intl.
  Symp. on Computer Architecture (Proc. ISCA-43)}}.
\newblock


\bibitem[\protect\citeauthoryear{Han, Mao, and Dally}{Han
  et~al\mbox{.}}{2016b}]%
        {han:iclr16:deep-compression}
\bibfield{author}{\bibinfo{person}{Song Han}, \bibinfo{person}{Huizi Mao},
  {and} \bibinfo{person}{William~J. Dally}.} \bibinfo{year}{2016}\natexlab{b}.
\newblock \showarticletitle{Deep Compression: Compressing Deep Neural Networks
  with Pruning, Trained Quantization, and Huffman Coding}. In
  \bibinfo{booktitle}{\emph{Proc. of the 5th Intl. Conf. on Learning
  Representationas (Proc. ICLR'16)}}.
\newblock


\bibitem[\protect\citeauthoryear{Hester, Peters, Yun, Peterson, Skinner, Golla,
  Storer, Hearndon, Freeman, Lord, Halter, Kotz, and Sorber}{Hester
  et~al\mbox{.}}{2016}]%
        {amulet}
\bibfield{author}{\bibinfo{person}{Josiah Hester}, \bibinfo{person}{Travis
  Peters}, \bibinfo{person}{Tianlong Yun}, \bibinfo{person}{Ronald Peterson},
  \bibinfo{person}{Joseph Skinner}, \bibinfo{person}{Bhargav Golla},
  \bibinfo{person}{Kevin Storer}, \bibinfo{person}{Steven Hearndon},
  \bibinfo{person}{Kevin Freeman}, \bibinfo{person}{Sarah Lord},
  \bibinfo{person}{Ryan Halter}, \bibinfo{person}{David Kotz}, {and}
  \bibinfo{person}{Jacob Sorber}.} \bibinfo{year}{2016}\natexlab{}.
\newblock \showarticletitle{Amulet: An Energy-Efficient, Multi-Application
  Wearable Platform}. In \bibinfo{booktitle}{\emph{Proceedings of the 14th ACM
  Conference on Embedded Network Sensor Systems}}
  \emph{(\bibinfo{series}{SenSys '16})}. \bibinfo{publisher}{ACM},
  \bibinfo{address}{New York, NY, USA}, \bibinfo{pages}{216--229}.
\newblock
\showISBNx{978-1-4503-4263-6}
\urldef\tempurl%
\url{https://doi.org/10.1145/2994551.2994554}
\showDOI{\tempurl}


\bibitem[\protect\citeauthoryear{Hester, Sitanayah, and Sorber}{Hester
  et~al\mbox{.}}{2015}]%
        {ufop}
\bibfield{author}{\bibinfo{person}{Josiah Hester}, \bibinfo{person}{Lanny
  Sitanayah}, {and} \bibinfo{person}{Jacob Sorber}.}
  \bibinfo{year}{2015}\natexlab{}.
\newblock \showarticletitle{Tragedy of the Coulombs: Federating Energy Storage
  for Tiny, Intermittently-Powered Sensors}. In
  \bibinfo{booktitle}{\emph{Proceedings of the 13th ACM Conference on Embedded
  Networked Sensor Systems}} \emph{(\bibinfo{series}{SenSys '15})}.
  \bibinfo{publisher}{ACM}, \bibinfo{address}{New York, NY, USA},
  \bibinfo{pages}{5--16}.
\newblock
\showISBNx{978-1-4503-3631-4}
\urldef\tempurl%
\url{https://doi.org/10.1145/2809695.2809707}
\showDOI{\tempurl}


\bibitem[\protect\citeauthoryear{Hester and Sorber}{Hester and Sorber}{[n.
  d.]}]%
        {flicker}
\bibfield{author}{\bibinfo{person}{Josiah Hester} {and} \bibinfo{person}{Jacob
  Sorber}.} \bibinfo{year}{[n. d.]}\natexlab{}.
\newblock \showarticletitle{Flicker: Rapid Prototyping for the Batteryless
  Internet of Things}. In \bibinfo{booktitle}{\emph{Proceedings of the 15th ACM
  Conference on Embedded Network Sensor Systems}}
  \emph{(\bibinfo{series}{SenSys '17})}.
\newblock


\bibitem[\protect\citeauthoryear{Hester, Storer, and Sorber}{Hester
  et~al\mbox{.}}{[n. d.]}]%
        {mayfly}
\bibfield{author}{\bibinfo{person}{Josiah Hester}, \bibinfo{person}{Kevin
  Storer}, {and} \bibinfo{person}{Jacob Sorber}.} \bibinfo{year}{[n.
  d.]}\natexlab{}.
\newblock \showarticletitle{Timely Execution on Intermi!ently Powered
  Ba!eryless Sensors}. In \bibinfo{booktitle}{\emph{Proceedings of the 15th ACM
  Conference on Embedded Network Sensor Systems}}
  \emph{(\bibinfo{series}{SenSys '17})}.
\newblock


\bibitem[\protect\citeauthoryear{Hicks}{Hicks}{2017}]%
        {clank}
\bibfield{author}{\bibinfo{person}{Matthew Hicks}.}
  \bibinfo{year}{2017}\natexlab{}.
\newblock \showarticletitle{Clank: Architectural Support for Intermittent
  Computation}. In \bibinfo{booktitle}{\emph{Proceedings of the 44th Annual
  International Symposium on Computer Architecture}}
  \emph{(\bibinfo{series}{ISCA '17})}. \bibinfo{publisher}{ACM},
  \bibinfo{address}{New York, NY, USA}, \bibinfo{pages}{228--240}.
\newblock
\showISBNx{978-1-4503-4892-8}
\urldef\tempurl%
\url{https://doi.org/10.1145/3079856.3080238}
\showDOI{\tempurl}


\bibitem[\protect\citeauthoryear{Horowitz}{Horowitz}{2014}]%
        {horowitz:isscc14:energy-keynote}
\bibfield{author}{\bibinfo{person}{Mark Horowitz}.}
  \bibinfo{year}{2014}\natexlab{}.
\newblock \showarticletitle{Computing's energy problem (and what we can do
  about it)}. In \bibinfo{booktitle}{\emph{Solid-State Circuits Conference
  Digest of Technical Papers (ISSCC), 2014 IEEE International}}. IEEE,
  \bibinfo{pages}{10--14}.
\newblock


\bibitem[\protect\citeauthoryear{Ignatov}{Ignatov}{[n. d.]}]%
        {har}
\bibfield{author}{\bibinfo{person}{Andrey Ignatov}.} \bibinfo{year}{[n.
  d.]}\natexlab{}.
\newblock \bibinfo{title}{HAR}.
\newblock
\newblock
\urldef\tempurl%
\url{https://github.com/aiff22/HAR}
\showURL{%
\tempurl}


\bibitem[\protect\citeauthoryear{Ioffe and Szegedy}{Ioffe and Szegedy}{2015}]%
        {ioffe2015batch}
\bibfield{author}{\bibinfo{person}{Sergey Ioffe} {and}
  \bibinfo{person}{Christian Szegedy}.} \bibinfo{year}{2015}\natexlab{}.
\newblock \showarticletitle{Batch normalization: Accelerating deep network
  training by reducing internal covariate shift}.
\newblock \bibinfo{journal}{\emph{arXiv preprint arXiv:1502.03167}}
  (\bibinfo{year}{2015}).
\newblock


\bibitem[\protect\citeauthoryear{Jayakumar, Raha, and Raghunathan}{Jayakumar
  et~al\mbox{.}}{2014}]%
        {quickrecall}
\bibfield{author}{\bibinfo{person}{H. Jayakumar}, \bibinfo{person}{A. Raha},
  {and} \bibinfo{person}{V. Raghunathan}.} \bibinfo{year}{2014}\natexlab{}.
\newblock \showarticletitle{{QuickRecall}: A Low Overhead {HW/SW} Approach for
  Enabling Computations across Power Cycles in Transiently Powered Computers}.
  In \bibinfo{booktitle}{\emph{Int'l Conf. on VLSI Design and Int'l Conf. on
  Embedded Systems}}.
\newblock


\bibitem[\protect\citeauthoryear{Jouppi, Young, Patil, Patterson, Agrawal,
  Bajwa, Bates, Bhatia, Boden, Borchers, et~al\mbox{.}}{Jouppi
  et~al\mbox{.}}{2017}]%
        {jouppi:isca17:tpu}
\bibfield{author}{\bibinfo{person}{Norman~P Jouppi}, \bibinfo{person}{Cliff
  Young}, \bibinfo{person}{Nishant Patil}, \bibinfo{person}{David Patterson},
  \bibinfo{person}{Gaurav Agrawal}, \bibinfo{person}{Raminder Bajwa},
  \bibinfo{person}{Sarah Bates}, \bibinfo{person}{Suresh Bhatia},
  \bibinfo{person}{Nan Boden}, \bibinfo{person}{Al Borchers}, {et~al\mbox{.}}}
  \bibinfo{year}{2017}\natexlab{}.
\newblock \showarticletitle{In-datacenter performance analysis of a tensor
  processing unit}.
\newblock \bibinfo{journal}{\emph{arXiv preprint arXiv:1704.04760}}
  (\bibinfo{year}{2017}).
\newblock


\bibitem[\protect\citeauthoryear{Krizhevsky, Sutskever, and Hinton}{Krizhevsky
  et~al\mbox{.}}{2012}]%
        {alexnet}
\bibfield{author}{\bibinfo{person}{Alex Krizhevsky}, \bibinfo{person}{Ilya
  Sutskever}, {and} \bibinfo{person}{Geoffrey~E Hinton}.}
  \bibinfo{year}{2012}\natexlab{}.
\newblock \showarticletitle{Imagenet classification with deep convolutional
  neural networks}. In \bibinfo{booktitle}{\emph{Advances in neural information
  processing systems}}. \bibinfo{pages}{1097--1105}.
\newblock


\bibitem[\protect\citeauthoryear{Kwon, Samajdar, and Krishna}{Kwon
  et~al\mbox{.}}{2018}]%
        {maeri}
\bibfield{author}{\bibinfo{person}{Hyoukjun Kwon}, \bibinfo{person}{Ananda
  Samajdar}, {and} \bibinfo{person}{Tushar Krishna}.}
  \bibinfo{year}{2018}\natexlab{}.
\newblock \showarticletitle{MAERI: Enabling Flexible Dataflow Mapping over DNN
  Accelerators via Reconfigurable Interconnects}. In
  \bibinfo{booktitle}{\emph{Proceedings of the Twenty-Third International
  Conference on Architectural Support for Programming Languages and Operating
  Systems}} \emph{(\bibinfo{series}{ASPLOS '18})}. \bibinfo{publisher}{ACM},
  \bibinfo{address}{New York, NY, USA}, \bibinfo{pages}{461--475}.
\newblock
\showISBNx{978-1-4503-4911-6}
\urldef\tempurl%
\url{https://doi.org/10.1145/3173162.3173176}
\showDOI{\tempurl}


\bibitem[\protect\citeauthoryear{Le~Cun, Jackel, Boser, Denker, Graf, Guyon,
  Henderson, Howard, and Hubbard}{Le~Cun et~al\mbox{.}}{1989}]%
        {lecun:ieee89:lenet}
\bibfield{author}{\bibinfo{person}{Yann Le~Cun}, \bibinfo{person}{LD Jackel},
  \bibinfo{person}{B Boser}, \bibinfo{person}{JS Denker}, \bibinfo{person}{HP
  Graf}, \bibinfo{person}{I Guyon}, \bibinfo{person}{D Henderson},
  \bibinfo{person}{RE Howard}, {and} \bibinfo{person}{W Hubbard}.}
  \bibinfo{year}{1989}\natexlab{}.
\newblock \showarticletitle{Handwritten digit recognition: Applications of
  neural network chips and automatic learning}.
\newblock \bibinfo{journal}{\emph{IEEE Communications Magazine}}
  \bibinfo{volume}{27}, \bibinfo{number}{11} (\bibinfo{year}{1989}),
  \bibinfo{pages}{41--46}.
\newblock


\bibitem[\protect\citeauthoryear{LeCun}{LeCun}{1998}]%
        {lecun1998mnist}
\bibfield{author}{\bibinfo{person}{Yann LeCun}.}
  \bibinfo{year}{1998}\natexlab{}.
\newblock \showarticletitle{The MNIST database of handwritten digits}.
\newblock \bibinfo{journal}{\emph{http://yann. lecun. com/exdb/mnist/}}
  (\bibinfo{year}{1998}).
\newblock


\bibitem[\protect\citeauthoryear{LeCun, Bottou, Bengio, and Haffner}{LeCun
  et~al\mbox{.}}{1998}]%
        {lecun1998gradient}
\bibfield{author}{\bibinfo{person}{Yann LeCun}, \bibinfo{person}{L{\'e}on
  Bottou}, \bibinfo{person}{Yoshua Bengio}, {and} \bibinfo{person}{Patrick
  Haffner}.} \bibinfo{year}{1998}\natexlab{}.
\newblock \showarticletitle{Gradient-based learning applied to document
  recognition}.
\newblock \bibinfo{journal}{\emph{Proc. IEEE}} \bibinfo{volume}{86},
  \bibinfo{number}{11} (\bibinfo{year}{1998}), \bibinfo{pages}{2278--2324}.
\newblock


\bibitem[\protect\citeauthoryear{Lucia and Ransford}{Lucia and
  Ransford}{2015}]%
        {dino}
\bibfield{author}{\bibinfo{person}{Brandon Lucia} {and}
  \bibinfo{person}{Benjamin Ransford}.} \bibinfo{year}{2015}\natexlab{}.
\newblock \showarticletitle{A Simpler, Safer Programming and Execution Model
  for Intermittent Systems}. In \bibinfo{booktitle}{\emph{Proceedings of the
  36th ACM SIGPLAN Conference on Programming Language Design and
  Implementation}} \emph{(\bibinfo{series}{PLDI 2015})}.
  \bibinfo{publisher}{ACM}, \bibinfo{address}{New York, NY, USA},
  \bibinfo{pages}{575--585}.
\newblock
\showISBNx{978-1-4503-3468-6}
\urldef\tempurl%
\url{https://doi.org/10.1145/2737924.2737978}
\showDOI{\tempurl}


\bibitem[\protect\citeauthoryear{Ma, Li, Li, Liu, Xie, Sampson, Kandemir, and
  Narayanan}{Ma et~al\mbox{.}}{2017}]%
        {ma2017incidental}
\bibfield{author}{\bibinfo{person}{Kaisheng Ma}, \bibinfo{person}{Xueqing Li},
  \bibinfo{person}{Jinyang Li}, \bibinfo{person}{Yongpan Liu},
  \bibinfo{person}{Yuan Xie}, \bibinfo{person}{Jack Sampson},
  \bibinfo{person}{Mahmut~Taylan Kandemir}, {and}
  \bibinfo{person}{Vijaykrishnan Narayanan}.} \bibinfo{year}{2017}\natexlab{}.
\newblock \showarticletitle{Incidental computing on IoT nonvolatile
  processors}. In \bibinfo{booktitle}{\emph{Proceedings of the 50th Annual
  IEEE/ACM International Symposium on Microarchitecture}}. ACM,
  \bibinfo{pages}{204--218}.
\newblock


\bibitem[\protect\citeauthoryear{Ma, Zheng, Li, Swaminathan, Li, Liu, Sampson,
  Xie, and Narayanan}{Ma et~al\mbox{.}}{2015}]%
        {nvp}
\bibfield{author}{\bibinfo{person}{Kaisheng Ma}, \bibinfo{person}{Yang Zheng},
  \bibinfo{person}{Shuangchen Li}, \bibinfo{person}{Karthik Swaminathan},
  \bibinfo{person}{Xueqing Li}, \bibinfo{person}{Yongpan Liu},
  \bibinfo{person}{Jack Sampson}, \bibinfo{person}{Yuan Xie}, {and}
  \bibinfo{person}{Vijaykrishnan Narayanan}.} \bibinfo{year}{2015}\natexlab{}.
\newblock \showarticletitle{Architecture exploration for ambient energy
  harvesting nonvolatile processors}. In \bibinfo{booktitle}{\emph{High
  Performance Computer Architecture (HPCA), 2015 IEEE 21st International
  Symposium on}}. IEEE, \bibinfo{pages}{526--537}.
\newblock


\bibitem[\protect\citeauthoryear{{Maeng}, {Colin}, and {Lucia}}{{Maeng}
  et~al\mbox{.}}{2017}]%
        {alpaca}
\bibfield{author}{\bibinfo{person}{Kiwan {Maeng}}, \bibinfo{person}{Alexei
  {Colin}}, {and} \bibinfo{person}{Brandon {Lucia}}.}
  \bibinfo{year}{2017}\natexlab{}.
\newblock \showarticletitle{Alpaca: Intermittent Execution without
  Checkpoints}. In \bibinfo{booktitle}{\emph{Proceedings of the ACM
  International Conference on Object Oriented Programming Systems Languages and
  Applications (OOPSLA)}}. \bibinfo{publisher}{ACM},
  \bibinfo{address}{Vancouver, BC, Canada}.
\newblock


\bibitem[\protect\citeauthoryear{Maeng and Lucia}{Maeng and Lucia}{2018}]%
        {maeng:osdi18:chinchilla}
\bibfield{author}{\bibinfo{person}{Kiwan Maeng} {and} \bibinfo{person}{Brandon
  Lucia}.} \bibinfo{year}{2018}\natexlab{}.
\newblock \showarticletitle{Adaptive Dynamic Checkpointing for Safe Efficient
  Intermittent Computing}. In \bibinfo{booktitle}{\emph{Proceedings of the 12th
  USENIX Conference on Operating Systems Design and Implementation}}
  \emph{(\bibinfo{series}{OSDI'18})}. \bibinfo{publisher}{USENIX Association},
  \bibinfo{address}{Berkeley, CA, USA}, \bibinfo{pages}{129--144}.
\newblock
\showISBNx{978-1-931971-47-8}
\urldef\tempurl%
\url{http://dl.acm.org/citation.cfm?id=3291168.3291178}
\showURL{%
\tempurl}


\bibitem[\protect\citeauthoryear{Miguel, Ganesan, Badr, and Jerger}{Miguel
  et~al\mbox{.}}{2018}]%
        {jerger2017ehmodel}
\bibfield{author}{\bibinfo{person}{J.~San Miguel}, \bibinfo{person}{K.
  Ganesan}, \bibinfo{person}{M. Badr}, {and} \bibinfo{person}{N.~E. Jerger}.}
  \bibinfo{year}{2018}\natexlab{}.
\newblock \showarticletitle{The EH Model: Analytical Exploration of
  Energy-Harvesting Architectures}.
\newblock \bibinfo{journal}{\emph{IEEE Computer Architecture Letters}}
  \bibinfo{volume}{17}, \bibinfo{number}{1} (\bibinfo{date}{Jan}
  \bibinfo{year}{2018}), \bibinfo{pages}{76--79}.
\newblock
\showISSN{1556-6056}
\urldef\tempurl%
\url{https://doi.org/10.1109/LCA.2017.2777834}
\showDOI{\tempurl}


\bibitem[\protect\citeauthoryear{Mirhoseini, Songhori, and
  Koushanfar}{Mirhoseini et~al\mbox{.}}{2013}]%
        {idetic}
\bibfield{author}{\bibinfo{person}{A. Mirhoseini}, \bibinfo{person}{E.~M.
  Songhori}, {and} \bibinfo{person}{F. Koushanfar}.}
  \bibinfo{year}{2013}\natexlab{}.
\newblock \showarticletitle{Idetic: A High-level Synthesis Approach for
  Enabling Long Computations on Transiently-powered {ASICs}}. In
  \bibinfo{booktitle}{\emph{IEEE Pervasive Computing and Communication
  Conference (PerCom)}}.
\newblock
\urldef\tempurl%
\url{http://aceslab.org/sites/default/files/Idetic.pdf}
\showURL{%
\tempurl}


\bibitem[\protect\citeauthoryear{Mitchell}{Mitchell}{1997}]%
        {Mitchell:1997:ML:541177}
\bibfield{author}{\bibinfo{person}{Thomas~M. Mitchell}.}
  \bibinfo{year}{1997}\natexlab{}.
\newblock \bibinfo{booktitle}{\emph{Machine Learning} (\bibinfo{edition}{1}
  ed.)}.
\newblock \bibinfo{publisher}{McGraw-Hill, Inc.}, \bibinfo{address}{New York,
  NY, USA}.
\newblock
\showISBNx{0070428077, 9780070428072}


\bibitem[\protect\citeauthoryear{Moritz, Nishihara, Wang, Tumanov, Liaw, Liang,
  Paul, Jordan, and Stoica}{Moritz et~al\mbox{.}}{2017}]%
        {moritz2017ray}
\bibfield{author}{\bibinfo{person}{Philipp Moritz}, \bibinfo{person}{Robert
  Nishihara}, \bibinfo{person}{Stephanie Wang}, \bibinfo{person}{Alexey
  Tumanov}, \bibinfo{person}{Richard Liaw}, \bibinfo{person}{Eric Liang},
  \bibinfo{person}{William Paul}, \bibinfo{person}{Michael~I Jordan}, {and}
  \bibinfo{person}{Ion Stoica}.} \bibinfo{year}{2017}\natexlab{}.
\newblock \showarticletitle{Ray: A Distributed Framework for Emerging AI
  Applications}.
\newblock \bibinfo{journal}{\emph{arXiv preprint arXiv:1712.05889}}
  (\bibinfo{year}{2017}).
\newblock


\bibitem[\protect\citeauthoryear{Nabhan and Zomaya}{Nabhan and Zomaya}{1994}]%
        {nabhan1994toward}
\bibfield{author}{\bibinfo{person}{Tarek~M Nabhan} {and}
  \bibinfo{person}{Albert~Y Zomaya}.} \bibinfo{year}{1994}\natexlab{}.
\newblock \showarticletitle{Toward generating neural network structures for
  function approximation}.
\newblock \bibinfo{journal}{\emph{Neural Networks}} \bibinfo{volume}{7},
  \bibinfo{number}{1} (\bibinfo{year}{1994}), \bibinfo{pages}{89--99}.
\newblock


\bibitem[\protect\citeauthoryear{Naderiparizi, Kapetanovic, and
  Smith}{Naderiparizi et~al\mbox{.}}{2016}]%
        {wispcam}
\bibfield{author}{\bibinfo{person}{Saman Naderiparizi}, \bibinfo{person}{Zerina
  Kapetanovic}, {and} \bibinfo{person}{Joshua~R. Smith}.}
  \bibinfo{year}{2016}\natexlab{}.
\newblock \showarticletitle{WISPCam: An RF-Powered Smart Camera for Machine
  Vision Applications}. In \bibinfo{booktitle}{\emph{Proceedings of the 4th
  International Workshop on Energy Harvesting and Energy-Neutral Sensing
  Systems}} \emph{(\bibinfo{series}{ENSsys'16})}. \bibinfo{publisher}{ACM},
  \bibinfo{address}{New York, NY, USA}, \bibinfo{pages}{19--22}.
\newblock
\showISBNx{978-1-4503-4532-3}
\urldef\tempurl%
\url{https://doi.org/10.1145/2996884.2996888}
\showDOI{\tempurl}


\bibitem[\protect\citeauthoryear{Nakkiran, Alvarez, Prabhavalkar, and
  Parada}{Nakkiran et~al\mbox{.}}{2015}]%
        {nakkiran:interspeech15:compressing}
\bibfield{author}{\bibinfo{person}{Preetum Nakkiran}, \bibinfo{person}{Raziel
  Alvarez}, \bibinfo{person}{Rohit Prabhavalkar}, {and}
  \bibinfo{person}{Carolina Parada}.} \bibinfo{year}{2015}\natexlab{}.
\newblock \showarticletitle{Compressing deep neural networks using a
  rank-constrained topology.}. In \bibinfo{booktitle}{\emph{Sixteenth Annual
  Conference of the International Speech Communication Association}}.
\newblock


\bibitem[\protect\citeauthoryear{Nowatzki, Gangadhar, and
  Sankaralingam}{Nowatzki et~al\mbox{.}}{2015}]%
        {nowatzki2015exploring}
\bibfield{author}{\bibinfo{person}{Tony Nowatzki}, \bibinfo{person}{Vinay
  Gangadhar}, {and} \bibinfo{person}{Karthikeyan Sankaralingam}.}
  \bibinfo{year}{2015}\natexlab{}.
\newblock \showarticletitle{Exploring the potential of heterogeneous von
  neumann/dataflow execution models}. In \bibinfo{booktitle}{\emph{ACM SIGARCH
  Computer Architecture News}}, Vol.~\bibinfo{volume}{43}. ACM,
  \bibinfo{pages}{298--310}.
\newblock


\bibitem[\protect\citeauthoryear{Parashar, Rhu, Mukkara, Puglielli, Venkatesan,
  Khailany, Emer, Keckler, and Dally}{Parashar et~al\mbox{.}}{2017}]%
        {parashar:isca17:scnn}
\bibfield{author}{\bibinfo{person}{Angshuman Parashar}, \bibinfo{person}{Minsoo
  Rhu}, \bibinfo{person}{Anurag Mukkara}, \bibinfo{person}{Antonio Puglielli},
  \bibinfo{person}{Rangharajan Venkatesan}, \bibinfo{person}{Brucek Khailany},
  \bibinfo{person}{Joel Emer}, \bibinfo{person}{Stephen~W. Keckler}, {and}
  \bibinfo{person}{William~J. Dally}.} \bibinfo{year}{2017}\natexlab{}.
\newblock \showarticletitle{SCNN: An Accelerator for Compressed-sparse
  Convolutional Neural Networks}. In \bibinfo{booktitle}{\emph{Proc. of the
  44th annual Intl. Symp. on Computer Architecture (Proc. ISCA-44)}}.
\newblock


\bibitem[\protect\citeauthoryear{Prabhakar, Zhang, Koeplinger, Feldman, Zhao,
  Hadjis, Pedram, Kozyrakis, and Olukotun}{Prabhakar et~al\mbox{.}}{2017}]%
        {plasticine}
\bibfield{author}{\bibinfo{person}{Raghu Prabhakar}, \bibinfo{person}{Yaqi
  Zhang}, \bibinfo{person}{David Koeplinger}, \bibinfo{person}{Matt Feldman},
  \bibinfo{person}{Tian Zhao}, \bibinfo{person}{Stefan Hadjis},
  \bibinfo{person}{Ardavan Pedram}, \bibinfo{person}{Christos Kozyrakis}, {and}
  \bibinfo{person}{Kunle Olukotun}.} \bibinfo{year}{2017}\natexlab{}.
\newblock \showarticletitle{Plasticine: A reconfigurable architecture for
  parallel patterns}. In \bibinfo{booktitle}{\emph{Computer Architecture
  (ISCA), 2017 ACM/IEEE 44th Annual International Symposium on}}. IEEE,
  \bibinfo{pages}{389--402}.
\newblock


\bibitem[\protect\citeauthoryear{Price, Glass, and Chandrakasan}{Price
  et~al\mbox{.}}{2018}]%
        {price2018speech}
\bibfield{author}{\bibinfo{person}{M. Price}, \bibinfo{person}{J. Glass}, {and}
  \bibinfo{person}{A.~P. Chandrakasan}.} \bibinfo{year}{2018}\natexlab{}.
\newblock \showarticletitle{A Low-Power Speech Recognizer and Voice Activity
  Detector Using Deep Neural Networks}.
\newblock \bibinfo{journal}{\emph{IEEE Journal of Solid-State Circuits}}
  \bibinfo{volume}{53}, \bibinfo{number}{1} (\bibinfo{date}{Jan}
  \bibinfo{year}{2018}), \bibinfo{pages}{66--75}.
\newblock
\showISSN{0018-9200}
\urldef\tempurl%
\url{https://doi.org/10.1109/JSSC.2017.2752838}
\showDOI{\tempurl}


\bibitem[\protect\citeauthoryear{Ransford, Sorber, and Fu}{Ransford
  et~al\mbox{.}}{2011}]%
        {mementos}
\bibfield{author}{\bibinfo{person}{Benjamin Ransford}, \bibinfo{person}{Jacob
  Sorber}, {and} \bibinfo{person}{Kevin Fu}.} \bibinfo{year}{2011}\natexlab{}.
\newblock \showarticletitle{Mementos: System Support for Long-Running
  Computation on {RFID}-Scale Devices}. In \bibinfo{booktitle}{\emph{ASPLOS}}.
\newblock


\bibitem[\protect\citeauthoryear{Ren, Li, Ding, Qiu, Wang, Li, Qian, and
  Yuan}{Ren et~al\mbox{.}}{2017}]%
        {ren2017sc}
\bibfield{author}{\bibinfo{person}{Ao Ren}, \bibinfo{person}{Zhe Li},
  \bibinfo{person}{Caiwen Ding}, \bibinfo{person}{Qinru Qiu},
  \bibinfo{person}{Yanzhi Wang}, \bibinfo{person}{Ji Li},
  \bibinfo{person}{Xuehai Qian}, {and} \bibinfo{person}{Bo Yuan}.}
  \bibinfo{year}{2017}\natexlab{}.
\newblock \showarticletitle{Sc-dcnn: highly-scalable deep convolutional neural
  network using stochastic computing}. In \bibinfo{booktitle}{\emph{Proceedings
  of the Twenty-Second International Conference on Architectural Support for
  Programming Languages and Operating Systems}}. ACM,
  \bibinfo{pages}{405--418}.
\newblock


\bibitem[\protect\citeauthoryear{Sainath and Parada}{Sainath and
  Parada}{2015}]%
        {okgoogle}
\bibfield{author}{\bibinfo{person}{Tara~N Sainath} {and}
  \bibinfo{person}{Carolina Parada}.} \bibinfo{year}{2015}\natexlab{}.
\newblock \showarticletitle{Convolutional neural networks for small-footprint
  keyword spotting}. In \bibinfo{booktitle}{\emph{16th Annual Conference of the
  International Speech Communication Association}}.
\newblock


\bibitem[\protect\citeauthoryear{Sample, Yeager, Powledge, Mamishev, and
  Smith}{Sample et~al\mbox{.}}{2008}]%
        {wisp}
\bibfield{author}{\bibinfo{person}{Alanson~P. Sample},
  \bibinfo{person}{Daniel~J. Yeager}, \bibinfo{person}{Pauline~S. Powledge},
  \bibinfo{person}{Alexander~V. Mamishev}, {and} \bibinfo{person}{Joshua~R.
  Smith}.} \bibinfo{year}{2008}\natexlab{}.
\newblock \showarticletitle{Design of an {RFID}-Based Battery-Free Programmable
  Sensing Platform}.
\newblock \bibinfo{journal}{\emph{IEEE Transactions on Instrumentation and
  Measurement}} \bibinfo{volume}{57}, \bibinfo{number}{11}
  (\bibinfo{date}{Nov.} \bibinfo{year}{2008}), \bibinfo{pages}{2608--2615}.
\newblock


\bibitem[\protect\citeauthoryear{Sankaralingam, Nagarajan, Liu, Kim, Huh,
  Burger, Keckler, and Moore}{Sankaralingam et~al\mbox{.}}{2003}]%
        {trips}
\bibfield{author}{\bibinfo{person}{Karthikeyan Sankaralingam},
  \bibinfo{person}{Ramadass Nagarajan}, \bibinfo{person}{Haiming Liu},
  \bibinfo{person}{Changkyu Kim}, \bibinfo{person}{Jaehyuk Huh},
  \bibinfo{person}{Doug Burger}, \bibinfo{person}{Stephen~W Keckler}, {and}
  \bibinfo{person}{Charles~R Moore}.} \bibinfo{year}{2003}\natexlab{}.
\newblock \showarticletitle{Exploiting ILP, TLP, and DLP with the polymorphous
  TRIPS architecture}. In \bibinfo{booktitle}{\emph{Computer Architecture,
  2003. Proceedings. 30th Annual International Symposium on}}. IEEE,
  \bibinfo{pages}{422--433}.
\newblock


\bibitem[\protect\citeauthoryear{Simonyan and Zisserman}{Simonyan and
  Zisserman}{2014}]%
        {vgg}
\bibfield{author}{\bibinfo{person}{Karen Simonyan} {and}
  \bibinfo{person}{Andrew Zisserman}.} \bibinfo{year}{2014}\natexlab{}.
\newblock \showarticletitle{Very deep convolutional networks for large-scale
  image recognition}.
\newblock \bibinfo{journal}{\emph{arXiv preprint arXiv:1409.1556}}
  (\bibinfo{year}{2014}).
\newblock


\bibitem[\protect\citeauthoryear{Song, Zhong, Zhang, Hu, Liu, Zhang, Wang, and
  Li}{Song et~al\mbox{.}}{2018}]%
        {song2018insitu}
\bibfield{author}{\bibinfo{person}{M. Song}, \bibinfo{person}{K. Zhong},
  \bibinfo{person}{J. Zhang}, \bibinfo{person}{Y. Hu}, \bibinfo{person}{D.
  Liu}, \bibinfo{person}{W. Zhang}, \bibinfo{person}{J. Wang}, {and}
  \bibinfo{person}{T. Li}.} \bibinfo{year}{2018}\natexlab{}.
\newblock \showarticletitle{In-Situ AI: Towards Autonomous and Incremental Deep
  Learning for IoT Systems}. In \bibinfo{booktitle}{\emph{2018 IEEE
  International Symposium on High Performance Computer Architecture (HPCA)}}.
  \bibinfo{pages}{92--103}.
\newblock
\urldef\tempurl%
\url{https://doi.org/10.1109/HPCA.2018.00018}
\showDOI{\tempurl}


\bibitem[\protect\citeauthoryear{Szegedy, Ioffe, Vanhoucke, and Alemi}{Szegedy
  et~al\mbox{.}}{2017}]%
        {szegedy2017inception}
\bibfield{author}{\bibinfo{person}{Christian Szegedy}, \bibinfo{person}{Sergey
  Ioffe}, \bibinfo{person}{Vincent Vanhoucke}, {and}
  \bibinfo{person}{Alexander~A Alemi}.} \bibinfo{year}{2017}\natexlab{}.
\newblock \showarticletitle{Inception-v4, inception-resnet and the impact of
  residual connections on learning.}. In \bibinfo{booktitle}{\emph{AAAI}},
  Vol.~\bibinfo{volume}{4}. \bibinfo{pages}{12}.
\newblock


\bibitem[\protect\citeauthoryear{Szegedy, Liu, Jia, Sermanet, Reed, Anguelov,
  Erhan, Vanhoucke, and Rabinovich}{Szegedy et~al\mbox{.}}{2015a}]%
        {googlenet}
\bibfield{author}{\bibinfo{person}{Christian Szegedy}, \bibinfo{person}{Wei
  Liu}, \bibinfo{person}{Yangqing Jia}, \bibinfo{person}{Pierre Sermanet},
  \bibinfo{person}{Scott Reed}, \bibinfo{person}{Dragomir Anguelov},
  \bibinfo{person}{Dumitru Erhan}, \bibinfo{person}{Vincent Vanhoucke}, {and}
  \bibinfo{person}{Andrew Rabinovich}.} \bibinfo{year}{2015}\natexlab{a}.
\newblock \showarticletitle{Going deeper with convolutions}. In
  \bibinfo{booktitle}{\emph{Proceedings of the IEEE conference on computer
  vision and pattern recognition}}. \bibinfo{pages}{1--9}.
\newblock


\bibitem[\protect\citeauthoryear{Szegedy, Liu, Jia, Sermanet, Reed, Anguelov,
  Erhan, Vanhoucke, Rabinovich, et~al\mbox{.}}{Szegedy et~al\mbox{.}}{2015b}]%
        {szegedy2015going}
\bibfield{author}{\bibinfo{person}{Christian Szegedy}, \bibinfo{person}{Wei
  Liu}, \bibinfo{person}{Yangqing Jia}, \bibinfo{person}{Pierre Sermanet},
  \bibinfo{person}{Scott Reed}, \bibinfo{person}{Dragomir Anguelov},
  \bibinfo{person}{Dumitru Erhan}, \bibinfo{person}{Vincent Vanhoucke},
  \bibinfo{person}{Andrew Rabinovich}, {et~al\mbox{.}}}
  \bibinfo{year}{2015}\natexlab{b}.
\newblock \showarticletitle{Going deeper with convolutions}. Cvpr.
\newblock


\bibitem[\protect\citeauthoryear{Szegedy, Vanhoucke, Ioffe, Shlens, and
  Wojna}{Szegedy et~al\mbox{.}}{2016}]%
        {szegedy2016rethinking}
\bibfield{author}{\bibinfo{person}{Christian Szegedy}, \bibinfo{person}{Vincent
  Vanhoucke}, \bibinfo{person}{Sergey Ioffe}, \bibinfo{person}{Jon Shlens},
  {and} \bibinfo{person}{Zbigniew Wojna}.} \bibinfo{year}{2016}\natexlab{}.
\newblock \showarticletitle{Rethinking the inception architecture for computer
  vision}. In \bibinfo{booktitle}{\emph{Proceedings of the IEEE Conference on
  Computer Vision and Pattern Recognition}}. \bibinfo{pages}{2818--2826}.
\newblock


\bibitem[\protect\citeauthoryear{{TI Inc.}}{{TI Inc.}}{2014}]%
        {wolverine}
\bibfield{author}{\bibinfo{person}{{TI Inc.}}} \bibinfo{year}{2014}\natexlab{}.
\newblock \bibinfo{title}{{Overview for MSP430FRxx FRAM}}.
\newblock \bibinfo{howpublished}{\url{http://ti.com/wolverine}}.
\newblock
\newblock
\shownote{Visited July 28, 2014.}


\bibitem[\protect\citeauthoryear{Tucker}{Tucker}{1966}]%
        {tucker1966some}
\bibfield{author}{\bibinfo{person}{Ledyard~R Tucker}.}
  \bibinfo{year}{1966}\natexlab{}.
\newblock \showarticletitle{Some mathematical notes on three-mode factor
  analysis}.
\newblock \bibinfo{journal}{\emph{Psychometrika}} \bibinfo{volume}{31},
  \bibinfo{number}{3} (\bibinfo{year}{1966}), \bibinfo{pages}{279--311}.
\newblock


\bibitem[\protect\citeauthoryear{Van Der~Woude and Hicks}{Van Der~Woude and
  Hicks}{2016}]%
        {ratchet}
\bibfield{author}{\bibinfo{person}{Joel Van Der~Woude} {and}
  \bibinfo{person}{Matthew Hicks}.} \bibinfo{year}{2016}\natexlab{}.
\newblock \showarticletitle{Intermittent computation without hardware support
  or programmer intervention}. In \bibinfo{booktitle}{\emph{Proceedings of
  OSDI'16: 12th USENIX Symposium on Operating Systems Design and
  Implementation}}. \bibinfo{pages}{17}.
\newblock


\bibitem[\protect\citeauthoryear{Zhang, Du, Zhang, Lan, Liu, Li, Guo, Chen, and
  Chen}{Zhang et~al\mbox{.}}{2016}]%
        {zhang2016cambricon}
\bibfield{author}{\bibinfo{person}{Shijin Zhang}, \bibinfo{person}{Zidong Du},
  \bibinfo{person}{Lei Zhang}, \bibinfo{person}{Huiying Lan},
  \bibinfo{person}{Shaoli Liu}, \bibinfo{person}{Ling Li}, \bibinfo{person}{Qi
  Guo}, \bibinfo{person}{Tianshi Chen}, {and} \bibinfo{person}{Yunji Chen}.}
  \bibinfo{year}{2016}\natexlab{}.
\newblock \showarticletitle{Cambricon-X: An accelerator for sparse neural
  networks}. In \bibinfo{booktitle}{\emph{Microarchitecture (MICRO), 2016 49th
  Annual IEEE/ACM International Symposium on}}. IEEE, \bibinfo{pages}{1--12}.
\newblock


\end{thebibliography}

\end{document}